\documentclass[aps,pre,twocolumn,superscriptaddress]{revtex4-1}  
\usepackage{graphicx,amsmath,epsfig,psfrag,verbatim,color,enumitem,float}
\usepackage[]{xcolor}  
\usepackage{subfigure}
\usepackage{epstopdf}
\usepackage{comment}
\usepackage[normalem]{ulem}  
\usepackage[export]{adjustbox}

\usepackage{rotating} 
\usepackage{tikz}
\usepackage{datetime} 

\newcommand{\mytitle}{Deep Learning Hamiltonians from \textcolor{black}{Disordered} Image Data in Quantum Materials}

\begin{document}
\usetikzlibrary{shapes.geometric, arrows}
\tikzstyle{startstop} = [rectangle, rounded corners, minimum width=3cm, minimum height=1cm,text centered, draw=black, fill=red!30]
\tikzstyle{io} = [trapezium, trapezium left angle=70, trapezium right angle=110, trapezium stretches=true, minimum width=3cm, minimum height=1cm, text centered, draw=black, fill=blue!30]
\tikzstyle{process} = [rectangle, minimum width=3cm, minimum height=1cm, text centered, draw=black, fill=orange!30]
\tikzstyle{decision} = [diamond, minimum width=3cm, minimum height=1cm, text centered, draw=black, fill=green!30]
\tikzstyle{arrow} = [thick,->,>=stealth]
\tikzstyle{doublearrow} = [thick,<->,>=stealth]

\title{\mytitle}

\author{S.~Basak}
\affiliation{Department of Physics and Astronomy, Purdue University, West Lafayette, IN 47907, USA}
\affiliation{Purdue Quantum Science and Engineering Institute, West Lafayette, IN 47907, USA}
\author{M.~Alzate Banguero}
\affiliation{Laboratoire de Physique et d'\'Etude des Mat\'eriaux, ESPCI Paris, PSL Universit\'e, CNRS, Sorbonne Universit\'e, 75005 Paris, France}
\author{L.~Burzawa}
\affiliation{Department of Computer Science, Purdue University, West Lafayette, IN 47907, USA}
\author{F.~Simmons}
\affiliation{Department of Physics and Astronomy, Purdue University, West Lafayette, IN 47907, USA}
\affiliation{Purdue Quantum Science and Engineering Institute, West Lafayette, IN 47907, USA}
\author{P.~Salev}
\affiliation{Department of Physics and Astronomy, University of Denver, Denver, Colorado 80208, USA}
\affiliation{Department of Physics and Center for Advanced Nanoscience, University of California-San Diego, La Jolla, California 92093, USA}
\author{L.~Aigouy}
\affiliation{Laboratoire de Physique et d'\'Etude des Mat\'eriaux, ESPCI Paris, PSL Universit\'e, CNRS, Sorbonne Universit\'e, 75005 Paris, France}
\author{M.~M. Qazilbash}
\affiliation{Department of Physics, College of William and Mary,
Williamsburg, VA 23187, USA}
\author{I.~K.~Schuller}
\affiliation{Department of Physics and Center for Advanced Nanoscience, University of California-San Diego, La Jolla, California 92093, USA}
\author{D.~N. Basov}
\affiliation{Department of Physics, Columbia University, New York, New York 10027, USA}
\author{A.~Zimmers}
\affiliation{Laboratoire de Physique et d'\'Etude des Mat\'eriaux, ESPCI Paris, PSL Universit\'e, CNRS, Sorbonne Universit\'e, 75005 Paris, France}
\author{E.~W.~Carlson}
\affiliation{Department of Physics and Astronomy, Purdue University, West Lafayette, IN 47907, USA}
\affiliation{Purdue Quantum Science and Engineering Institute, West Lafayette, IN 47907, USA}

\date{\today}

\begin{abstract}
The capabilities of image probe experiments are rapidly expanding, providing new information about quantum materials on unprecedented length and time scales. Many such materials feature inhomogeneous electronic properties
with intricate pattern formation on the observable surface. This rich spatial structure contains information about interactions, dimensionality, and disorder -- a spatial encoding of the Hamiltonian driving the pattern formation. Image recognition techniques from machine learning are an excellent tool for interpreting information encoded in the spatial relationships in such images. Here, we develop a deep learning framework for using the rich information available in these spatial correlations in order to discover the underlying Hamiltonian driving the patterns.  We first vet the method on a known case, 
scanning near-field optical microscopy on a thin film of VO$_2$.  
We then apply our trained convolutional neural network architecture 
to new optical microscope images of a different VO$_2$ film as it goes through the metal-insulator transition. 
We find that a two-dimensional Hamiltonian
with both interactions and random field disorder
is required to explain the intricate, fractal intertwining of metal and insulator domains during the transition.  
This detailed knowledge about the underlying Hamiltonian paves the way to using the model to control the pattern formation
via, {\em e.g.}, tailored hysteresis protocols. 
We also introduce a distribution-based confidence measure on the results of a multi-label classifier, which does not rely on adversarial training.
In addition, we propose a new machine learning based criterion for diagnosing a physical system's proximity to criticality. 
\end{abstract}

\maketitle


\section{Introduction}
\label{sxn:introduction}

The types of surface probes, such as atomic force microscopy (AFM), scanning tunneling microscopy (STM), and scattering scanning near-field infrared Microscope (SNIM), among many others~\cite{RevModPhys.83.471,RevModPhys.71.S324} and the wealth of data they generate is increasing at a rapid pace.
As often happens in science, new experimental frontiers reveal new physics: These scanning and image probe experiments often reveal complex 
\textcolor{black}{
electronic pattern formation
spanning multiple length scales
at the surface of correlated 
quantum materials, even when they
are atomicaly smooth.}\cite{qazilbash-science,kohsaka-science,schuller-breakdown,schuller-filament,phillabaum-natcomm,shuo-prl,dagotto-science,moreo-manganites,allan-pnictide,allan-iridate,milan-iridates}
\textcolor{black}{
For example, manganites can have  ferromagnetic and antiferromagnetic regions that coexist on multiple length scales.\cite{dagotto-science}  
In the unidirectional electronic glass
in cuprates,\cite{kohsaka-science} domains of stripe orientation take fractal form with correlations over 4 orders of magnitude in length scale.\cite{phillabaum-2012}  Magnetic domains in NdNiO$_3$ were also revealed to have fractal textures.\cite{comin-ndnio}  We focus here on VO$_2$, a material whose metal and insulator domains
can show self-similar structure over multiple length scales.\textcolor{black}{\cite{qazilbash-science,shuo-prl,sharoni-avalanches}}
}

Unfortunately, most of our theoretical  tools are designed for understanding and describing homogeneous electronic states. Therefore it is vital that we envision new theoretical frameworks for understanding why the patterns form in strongly correlated materials.  The cluster analysis techniques we developed for interpreting these images have already uncovered universal behavior among disparate quantum materials,\cite{qazilbash-science,post-natphys,comin-ndnio,phillabaum-natcomm,shuo-prl} 
but the methods only work on systems near criticality, and for sufficiently large fields of view.  Powerful image recognition methods from machine learning (ML) hold potential to complement and extend these analyses into new regimes.

There has been tremendous growth recently in the application of ML methods to condensed matter.  
(For reviews, see Refs.~\onlinecite{carleo-ml-review-2019,carrasquilla-ml-review-2020,bedolla-ml-review-2021,mehta-review-2019}.)
ML is being applied as a tool to tackle various problems in 
condensed matter physics, including
disordered and glassy systems systems\cite{nussinov-ml-2011,schoenholz-ml,millis-prb},
quantum many body problems\cite{troyer-ml},  quantum transport\cite{lopez-ml}, renormalization group\cite{ML-prec4,ML-prec5}, and big data in
materials science\cite{schmidt-ml-review-2019,ghiringhelli-ml}. 
ML also benefits from physics, an area known as physics-inspired ML theory.\cite{carleo-ml-review-2019}
Applied to experimental data, ML has been used to detect which
phase of matter a physical system is in,\cite{kim-davis-ml-2020}
and aid in the experimental detection of the glass transition temperature.\cite{ml-glass-transition}  Other common uses
of ML for experimental data include the extraction 
of material parameters from experiment, \cite{schmidt-ml-review-2019,basov-ml-polariton}
or using ML to replace a lengthy and time consuming fitting procedure.\cite{basov-liu-ml-2021,basov-ml-polariton}

Regarding phase transitions, ML has been used to detect which phase of matter a theoretical configuration is in,\cite{mehta-review-2019,Carrasquilla2017,bedolla-ml-review-2021}
as well as identify the transition temperature of a theoretical model,\cite{Carrasquilla2017,bedolla-ml-review-2021,wang-ml}
each in cases where a particular Hamiltonian is already assumed.
Relatively little attention has been paid to the critical region,\cite{mehta-review-2019}  \textcolor{black}{where domains
display power law structure across multiple length scales.}
In addition, much of the work done to identify phases or detect phase transitions has been purely computational, with the Hamiltonian 
assumed.\cite{mehta-review-2019,Carrasquilla2017,bedolla-ml-review-2021,wang-ml}
By contrast, our method utilizes the rich spatial correlations available in near critical configurations to detect which {\em Hamiltonian} should be used to describe a physical system, and we apply the method to experimentally derived data.


Here we develop a Deep Learning (DL) classifier to recognize spatial configurations
from several different Hamiltonians.   We test the DL classifier  on 
experimental image data of VO$_2$ obtained via SNIM, and 
then apply it to new optical microscope images  of VO$_2$. 
Convolutional neural networks in particular are heavily used in image classification.
We have previously  shown that with ML, images from simulation can be classified with very good accuracy of $\sim97\%$.~\cite{lukasz-ML}
Here we show that a DL architecture can classify 2D surface images into one of seven candidate theoretical models,
to even better accuracy ($>99\%$). 
We introduce a symmetry reduction method
which reduces training time over the 
data augmentation method. 
In addition, we use the DL model on experimental images derived from SNIM and optical microscope data to discover the underlying Hamiltonian driving pattern
formation of metal and insulator puddles in films of VO$_2$.
We also introduce a new method for judging the confidence of a
multi-label classifier, based on the  
\textcolor{black}{multivariate distribution of values of the output nodes.}
We furthermore propose that this confidence
measure tracks proximity to criticality.



This article is organized as follows:
In Sec.~\ref{sxn:Simulations}, we give an overview of the Hamiltonians 
\textcolor{black}{
of interest
from statistical mechanics 
}
in this paper. 
Section~\ref{sxn:Customized_Deep_Learning_Model} shows the end-to-end deep learning architecture and process.
We demonstrate the effectiveness of 
\textcolor{black}{
symmetry reduction to reduce training time
as compared with data augmentation, 
}
and develop a confidence criterion to judge
the reliability of predictions.
In Sec.~\ref{sxn:Application_to_Experimental_Images}, we make predictions on SNIM and optical microscope data on 
thin films of VO$_2$.
We show that using only simulated data for training, we have developed a robust deep learning classification model, that can learn the Hamiltonian driving pattern formation from experimental surface probe images. 



\section{Developing a Deep Learning Model to Reveal Underlying Hamiltonians}

\textcolor{black}{
}

We first construct several possible Hamiltonians that could potentially describe the morphology of these metal and insulator domains,
including the multiscale behavior. 
Then, we use numerical simulations to generate thousands of spatial configurations of metal and insulator domains that can arise in these Hamiltonians.  Next, we 
develop and train a Deep Learning (DL) convolutional neural network 
on a subset of these images in supervised learning mode.  
After we validate that the DL model can correctly identify
the underlying Hamiltonian from a single domain configuration
with greater than 99\% accuracy, we then apply our trained
DL model to experimental data on VO$_2$ obtained via both
SNIM and optical microscopy.


\subsection{Candidate Hamiltonians and the Morphologies They Produce}
\label{sxn:Simulations}

\begin{figure*}
    \begin{center}
    \includegraphics[width=0.175\textwidth]{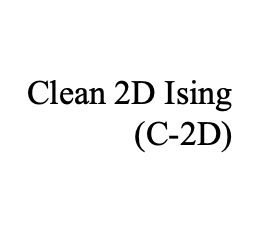}
    \subfigure[\ T=2.30J]{%
    \includegraphics[width=0.175\textwidth,frame]{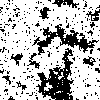}}
    \subfigure[\ T=2.35J]{%
    \includegraphics[width=0.175\textwidth,frame]{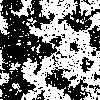}}
    \subfigure[\ T=2.45J]{%
    \includegraphics[width=0.175\textwidth,frame]{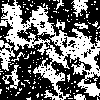}}
    \subfigure[\ T=2.55J]{%
    \includegraphics[width=0.175\textwidth,frame]{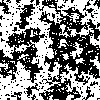}}\linebreak
    \includegraphics[width=0.175\textwidth]{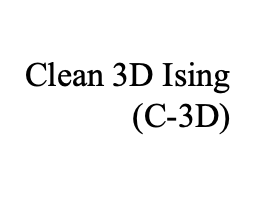}
    \subfigure[\ T=4.50J]{%
    \includegraphics[width=0.175\textwidth,frame]{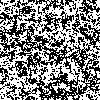}}
    \subfigure[\ T=4.52J]{%
    \includegraphics[width=0.175\textwidth,frame]{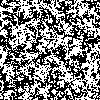}}
    \subfigure[\ T=4.55J]{%
    \includegraphics[width=0.175\textwidth,frame]{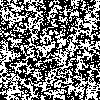}}
    \subfigure[\ T=4.60J]{%
    \includegraphics[width=0.175\textwidth,frame]{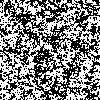}}\linebreak
    \includegraphics[width=0.175\textwidth]{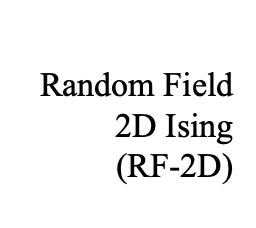}
    \subfigure[\ R=1.00J]{%
    \includegraphics[width=0.175\textwidth,frame]{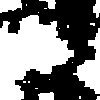}}
    \subfigure[\ R=1.05J]{%
    \includegraphics[width=0.175\textwidth,frame]{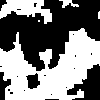}}
    \subfigure[\ R=1.10J]{%
    \includegraphics[width=0.175\textwidth,frame]{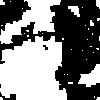}}
    \subfigure[\ R=1.15J]{%
    \includegraphics[width=0.175\textwidth,frame]{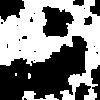}}\linebreak
    \includegraphics[width=0.175\textwidth]{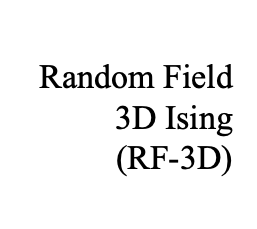}
    \subfigure[\ R=2.25J]{%
    \includegraphics[width=0.175\textwidth,frame]{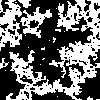}}
    \subfigure[\ R=2.26J]{%
    \includegraphics[width=0.175\textwidth,frame]{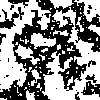}}
    \subfigure[\ R=2.27J]{%
    \includegraphics[width=0.175\textwidth,frame]{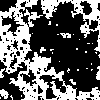}}
    \subfigure[\ R=2.28J]{%
    \includegraphics[width=0.175\textwidth,frame]{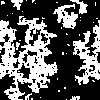}}\linebreak
    \includegraphics[width=0.175\textwidth]{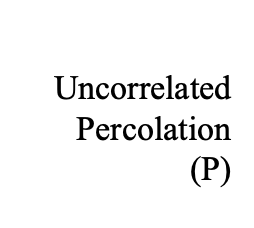}
    \subfigure[\ p=0.31]{%
    \includegraphics[width=0.175\textwidth,frame]{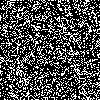}}
    \subfigure[\ p=0.50]{%
    \includegraphics[width=0.175\textwidth,frame]{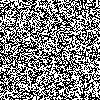}}
    \subfigure[\ p=0.59]{%
    \includegraphics[width=0.175\textwidth,frame]{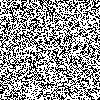}}
    \subfigure[\ p=0.80]{%
    \includegraphics[width=0.175\textwidth,frame]{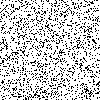}}
    \caption{%
    \label{fig:sample_configs}
        Typical critical configurations generated from simulations of clean and random field Ising models and percolation models.
    }
    \end{center}
\end{figure*}

We use numerical simulations to generate typical configurations of metal and insulator domains that can arise from various model Hamiltonians that could potentially be controlling the metal-insulator domain structure.
Simulation methods are described briefly in this section and in detail in the SI,
and are summarized in Table~\ref{tab:sim_params}.
As VO$_2$ undergoes a temperature-driven transition from metal to insulator and vice versa,
the macroscopic resistivity 
changes by 4 to 5 orders of magnitude.  
However, rather than doing so homogeneously,
\textcolor{black}{we previously}
used 
SNIM to produce spatially resolved images of the
metal and insulator domains which
revealed that VO$_2$ thin films transition {\em inhomogeneously},\footnote{\textcolor{black}{Note that the 
voltage-driven transition is also inhomogeneous, as revealed by optical measurements.\cite{schuller-quantumsensing}}}
with metal and insulator domains interleaving with each
other over a wide range of 
length scales.\cite{qazilbash-science,sohn-fractal}
We introduce a range of possible Hamiltonians which 
could be responsible for driving the multiscale textures
during the metal-insulator transition in VO$_2$.
\textcolor{black}{Domain}
configurations from these Hamiltonians will be 
used to train the DL model to identify the underlying
physics driving pattern formation in this material.

\textcolor{black}{
Because the experimental probes of interest are directly measuring electronic degrees of freedom, we construct Hamiltonians that are about these electronic degrees of freedom. 
The intricate patterns of metal and insulator domains happens across multiple length scales from the resolution of the probes all the way out to the field of view ($\approx 20nm - 4\mu m$ for SNIM and $\approx 370nm - 28 \mu m$ for optical microscope data).
Therefore, we  construct Hamiltonians at the order parameter level.
Because the theories are constructed at the order parameter level, they are not microscopic, although they can provide
constraints on microscopic models.  
}

\begin{figure}
    \begin{center}
    \includegraphics[width=\columnwidth]{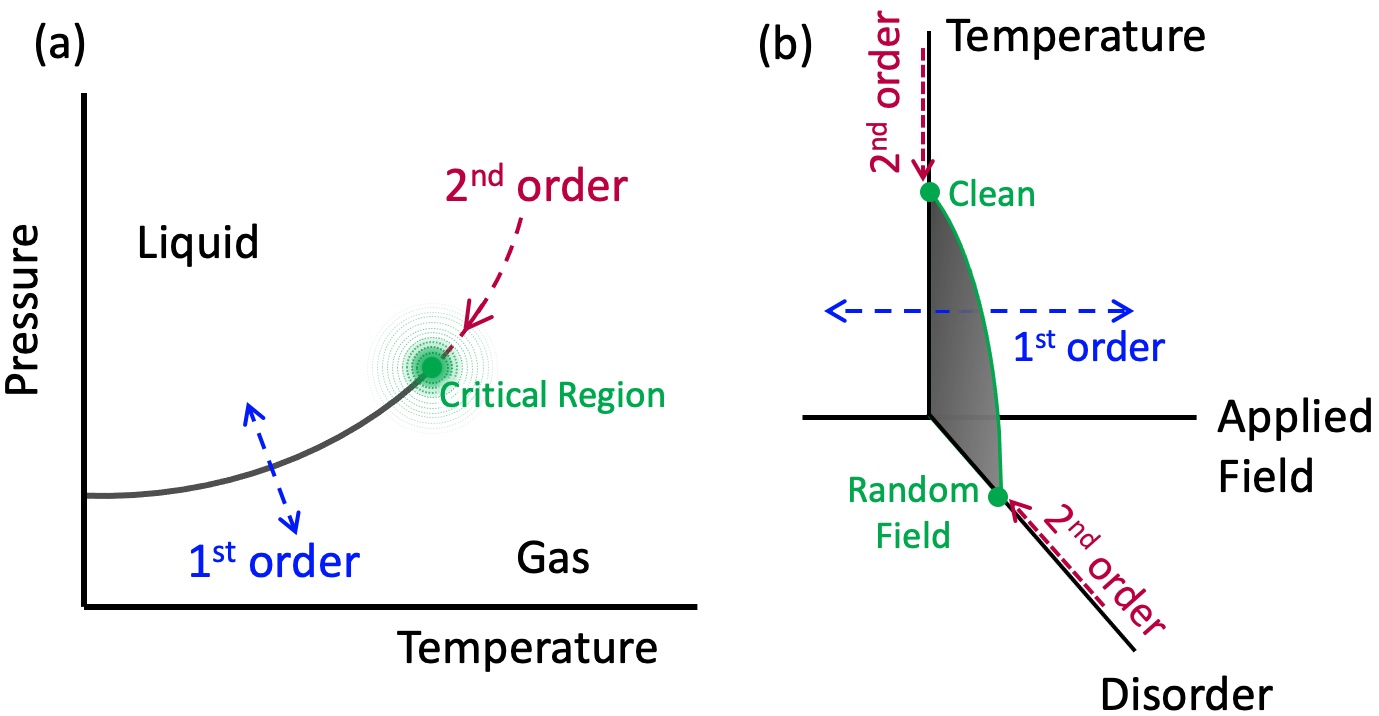}
    \caption{
    \textcolor{black}{
    (a) Generic liquid-gas phase diagram above the triple point.  For paths along the dotted blue line, the phase transition is first order.  However, the transition is second order when approached along the dotted red line.  The critical end point (solid green circle) exerts itself over a critical region (open green circles).  (b) Phase diagram of the random field Ising model.  While the clean or random field Ising model has a second order phase transition as a function of temperature (the red dotted lines), the transition is first order when approached as a function of applied field, crossing through the ordered region (blue dotted line).  Critical behavior is observed in the vicinity of the green critical temperature line, whose critical behavior is controlled by the random field fixed point for any finite disorder strength. Because the random field fixed point is a zero temperature fixed point, the critical region is much broader than in the clean case.
    } 
    }
        \label{fig:phasediagram}
    \end{center}
\end{figure}

First, we consider a clean, interacting Hamiltonian.  
A reasonable ansatz is that the interaction energy
between neighboring domains is lower for like domains
than for unlike domains.  We model this proclivity toward
neighboring like domains\cite{fradkin-phillips,shuo-prl} with a 
nearest-neighbor Ising Hamiltonian:  
\begin{equation}
    H = - J \sum_{<i,j>} \sigma_i \sigma_j - \sum_i h \sigma_i
\label{eqn:Ising_model} 
\end{equation}
where, $\sigma_i = \pm 1$
is a two-state local order parameter which in this case 
tracks metal ({\em e.g.}, $\sigma = 1$) and insulator ({\em e.g.}, $\sigma = -1$)
domains. 
In an infinite size system, this model undergoes 
\textcolor{black}{an equilibrium, second order}
phase transition 
\textcolor{black}{as a function of temperature}
at a critical temperature of $T_c^{2D}\approx2.27J$ in two-dimensional systems and $T_c^{3D}\approx4.51J$ in three-dimensional systems.~\cite{PhysRev.65.117,RevModPhys.46.597}
\textcolor{black}{
However, the model also undergoes a first order phase transition as a function of applied field $h$.  This first order line terminates in the critical
end point mentioned above.  The phenomenology of a first order
line terminating in a critical end point is why this model is 
often used in conjunction with the liquid-gas transition.
For example, when the liquid-gas transition is approached along
the coexistence curve in temperature and pressure, the
transition is second order (see the red dotted line in Fig.~\ref{fig:phasediagram}(a)).\cite{kittel-text}  
The influence of that critical point is felt throughout a 
critical region
(the light green region in Fig.~\ref{fig:phasediagram}(a)), which includes part of the first order line in the 
vicinity of the critical 
end point.
Similarly, this model can be used to
describe the first order metal-insulator transition in 
VO$_2$,
with a critical end point whose 
influence extends along the first order line.\cite{shuo-prl}  
The physical structure of domains is power law 
throughout this critical region, when viewed on length 
scales shorter than the correlation length, which diverges
as the critical point is approached.
In mapping this order parameter model to the temperature-driven
metal-insulator transition in VO$_2$, we are making the ansatz that
a sweep of temperature in the experiment maps to a combination
of temperature and field sweep in the model, as in 
our prior work\cite{shuo-prl} and Ref.~\cite{limelette}
} 

\begin{table*}
  \centering 
 \begin{tabular*}{\textwidth}{@{\extracolsep{\stretch{1}}}*{3}{c}@{}}
    \hline
    Model &  {Parameters Simulated} & {Simulation Method}\\ 
    \hline \hline 
    {2D Clean Ising (C-2D)}  & T = $2.25-2.64$  & Monte Carlo \\
    {3D Clean Ising (C-3D)} & T = $4.45-4.65$  & Monte Carlo \\
    {2D RFIM (RF-2D)}   & R = $1.00-1.19$  & Zero Temperature Field Sweep  \\
    {3D RFIM (RF-3D)}   & R = $2.25-2.29$  & Zero Temperature Field Sweep  \\
    {2D Percolation (P-2D)}   & p = $0.57-0.61$ & Biased coin flip \\
    {3D Percolation (P-3D)}   & p = $0.29-0.33$ & Biased coin flip \\
    \hline
                          &  p = $0.02-0.2$ & Biased coin flip \\
    {Non-Critical Percolation (P$^*$)}  & p = $0.48-0.52$ & Biased coin flip \\
                           & p = $0.8-0.98$ & Biased coin flip \\
    \hline
  \end{tabular*}
  \caption{Parameters of simulations of the statistical mechanics models.  In the first part of the table, parameters are in the critical region.  In the second part of the table, parameters are not near criticality.}
\label{tab:sim_params}
\end{table*}


We simulate configurations near criticality (see Table~\ref{tab:sim_params}),
since that is
where this Hamiltonian can cause structure over multiple length scales.  
We use Monte Carlo simulations to generate typical examples of multiscale morphologies of insulator and metal domains that can arise from 
\textcolor{black}{
the clean Ising Hamiltonians of Eqn.~\ref{eqn:Ising_model}.
}   
Intricate 
\textcolor{black}{domain configurations}
arise near the critical points of this model.  
Figure~\ref{fig:sample_configs}(a-d) shows some configurations near $T_c^{2D}$ on a $100\times100$ lattice, with periodic boundary conditions.
Figure~\ref{fig:sample_configs}(e-h) shows some representative configurations near  $T_c^{3D}$ on a $100\times100\times100$ lattice.
Further simulation details are in the SI.

The correlation length of a system diverges at criticality,
$\xi \propto 1/|T-T_c|^{\nu}$.  
When viewed on length scales $x < \xi$, the system exhibits
critical fluctuations, {\em i.e.} fluctuations on all length scales between the correlation length $\xi$ and the short distance cutoff, which for the lattice models we study is the lattice spacing, and in the real physical system it is the 
size of a unit cell. 
Close enough to criticality, this length scale will exceed any finite field of view (FOV).  Therefore, when observed on a finite
FOV (experimentally, or in simulation), there is a finite
range of parameters over which the system displays critical pattern formation.  For this reason, the entire range of parameters listed in the first part of 
Table~\ref{tab:sim_params} should be viewed as critical for the FOV's considered in this paper.

In addition to an interaction energy between domains,
material disorder also affects the types of shapes that metal
and insulator domains take.  
Because material disorder may make certain regions of the sample
more favorable to insulator, and certain others more favorable to metal,
we use a random field Ising model (RFIM) to 
simulate the effects of material disorder on the metal and insulator 
textures:\footnote{Disorder can also cause spatial variations in the coupling J. This random bond disorder is irrelevant in the renormalization group sense when random
field disorder is present.}
\begin{equation}
    H = - J \sum_{<ij>} \sigma_i \sigma_j - \sum_i (h_i + h) \sigma_i
\label{eqn:RFIM}
\end{equation}
The first term is the clean Ising model of  Eqn.~\ref{eqn:Ising_model}.
In the second term, the uniform field $h$ and the local random fields $h_i$ couple directly with the local order parameter. 
The random fields are chosen from a Gaussian distribution of width $R$ where the probability of $h_i$ is $P(h_i)= {\exp(-h_i^2/(2R^2))}/\sqrt{2\pi \sigma^2}$. 
In the physical system, VO$_2$ changes from insulator to metal
as the temperature is changed.  Within the model, this physics
presents itself as a combination of model temperature 
{\em and uniform field $h$}.\cite{shuo-prl}

\textcolor{black}{
The ordered phase corresponds to all metal or all insulator,
and the transition is second order when approached 
as a function of temperature or disorder strength at zero applied field (see the red dotted lines in Fig.~\ref{fig:phasediagram}(b)). 
When instead the field is swept across the ordered region, the transition is a first order change from metal to insulator
(see the blue dotted line in Fig.~\ref{fig:phasediagram}(b)).
When temperature and the disorder strength
are both nonzero,
}
the behavior of the model in the vicinity
of the phase transition is dominated by the random field.\cite{dan-fisher-rfim}
That is, the random field is relevant but the temperature is irrelevant
in the renormalization group sense
\textcolor{black}{
in a broad range around the solid green line in
Fig.~\ref{fig:phasediagram}(b).
} 
We therefore model the patterns
of metal and insulator domains that are possible with this Hamiltonian
by generating \textcolor{black}{domain configurations}
at zero temperature, while sweeping the uniform field $h$. 
\textcolor{black}{At zero temperature, this model undergoes
an equilibrium phase transition at a random field strength of 
$R_c \approx 2.27J$
in an infinite size three-dimensional system (RF-3D).\cite{middleton-fisher}
}
In two dimensions (RF-2D), the critical disorder strength is $R_c \rightarrow0$~\cite{PhysRevLett.75.4528} in the infinite size limit,
although in a finite size system or with finite FOV, $R_c(L) > 0$.  For the FOV we consider, $R_c \approx J$ for RF-2D.  

When the random field model is near criticality, as the uniform field is swept from low to high or high to low, 
intricate patterns develop over multiple length scales near the coercive field strength,
where the 
\textcolor{black}{metal/insulator domain fraction}
changes most rapidly with respect to uniform field $h$.
Figures~\ref{fig:sample_configs}(i-l) show representative configurations of RF-2D for a  $100\times100$ lattice. 
Figures~\ref{fig:sample_configs}(m-p) show representative configurations  on the surface of a $100\times100\times100$ lattice near the 3D critical
disorder strength, $R_c^{3D}$.

There is also the possibility that in fact domains are {\em not} interacting with each other
as in the above Hamiltonians, but rather each domain acts independently.  
In the corresponding uncorrelated percolation model, a site is labeled ``metallic'' with a probability $p$; otherwise it is labeled ``insulating''.
When $p \ne 0.5$, this is like flipping a biased coin where $p$ is the probability of turning up heads. 
This model also has a second order phase transition as a function of $p$,
and displays structure across multiple length scales near its
critical point. 
The critical percolation strength $p_c$ is marked by a percolating cluster spanning the entire system, meaning that it touches one side of the  system, and also the opposite side.  
In a two-dimensional system on an infinite square lattice, this threshold occurs at $p_c^{2D}\approx 0.59$ and in a three-dimensional system on an infinite cubic lattice it occurs at $p_c^{3D}\approx 0.31$.~\cite{PhysRevE.72.016126}
Figure~\ref{fig:sample_configs}(s) shows a percolation configuration of size $100\times100$ at  $p_c^{2D} = 0.59$. 
Figure~\ref{fig:sample_configs}(q) shows a percolation configuration $100\times100\times100$ at  $p_c^{3D} = 0.31$. 


In order to further train the DL model to distinguish
configurations that are near criticality (such as those described above)
from configurations that are not near criticality, 
we also generate training images on 
uncorrelated percolation away from any critical point. 
In order to avoid the  
multiscale, fractal textures associated with criticality, in this set of images
we use the percolation model in the following ranges:
$p = 0.02-0.2;~ 0.48-0.52;~0.8-0.98$.
The first range produces images which are mostly black;
the second range produces images which are ``white noise''
(such as Fig.~\ref{fig:sample_configs}(r)),
and the third range produces images which are mostly all white
(like those in Fig.~\ref{fig:sample_configs}(t)).
 Table~\ref{tab:sim_params} summarizes the parameter ranges
 we use for generating simulated data  for training and validation from each of the above Hamiltanions.


\section{Customized Deep Learning Model}
\label{sxn:Customized_Deep_Learning_Model}

\begin{figure}
    \centering
    \resizebox{.45\textwidth}{!}{%
    \begin{tikzpicture}[node distance=2cm]
        \node (in1) [io, anchor=center] {Image: $(100\times100\times1)$};
        \node (dec1) [decision, below of=in1, anchor=east, aspect=2, xshift=-1cm] {\textcolor{black}{Mostly metal?}};
        \node (pro2) [process, right of=dec1, anchor=west, xshift=2cm] {\textcolor{black}{Flip $\sigma$ $0\leftrightarrow1$}};
        \node (dec2) [decision, below of=dec1, aspect=2, xshift=0cm, yshift=-1.4cm] {\textcolor{black}{Quadrant with most metal?}};
        \node (pro3a) [process, right of=dec2, anchor=west, xshift=2cm, yshift=1.50cm] {Rotate $90^o$};
        \node (pro3b) [process, right of=dec2, anchor=west, xshift=2cm, yshift=0.00cm] {Rotate $90^o$};
        \node (pro3c) [process, right of=dec2, anchor=west, xshift=2cm, yshift=-1.5cm] {Rotate $90^o$};
        \node (dec3) [decision, below of=dec2, aspect=2, yshift=-1.4cm] {\textcolor{black}{More metal in 2nd than 4th quadrant?}};
        \node (pro4) [process, right of=dec3, anchor=west, xshift=2cm] {Transpose};
        \node (out1) [io, below of=dec3, xshift=2.5cm] {Output: $100\times100\times1$};
        \draw [arrow] (in1) -- (dec1);
        \draw [arrow] (dec1) -- node[anchor=west] {Yes} (dec2);
        \draw [arrow] (dec1) -- node[anchor=south] {No} (pro2);
        \draw [arrow] (pro2) -- (dec2);
        \draw [arrow] (dec2) -- node[anchor=north] {4th} (pro3a);
        \draw [arrow] (dec2) -- node[anchor=north] {3rd} (pro3b);
        \draw [arrow] (dec2) -- node[anchor=north] {2nd} (pro3c);
        \draw [arrow] (dec2) -- node[anchor=east] {1st} (dec3);
        \draw [arrow] (pro3a) -- (pro3b);
        \draw [arrow] (pro3b) -- (pro3c);
        \draw [arrow] (pro3c) -- (dec3);
        \draw [arrow] (dec3) -- node[anchor=west] {Yes} (out1);
        \draw [arrow] (dec3) -- node[anchor=south] {No} (pro4);
        \draw [arrow] (pro4) -- (out1);
    \end{tikzpicture}
    }
    \caption{%
    \label{fig:symmetry_CNN}
        Symmetry reduction method, as described in the text.  
    }
\end{figure}
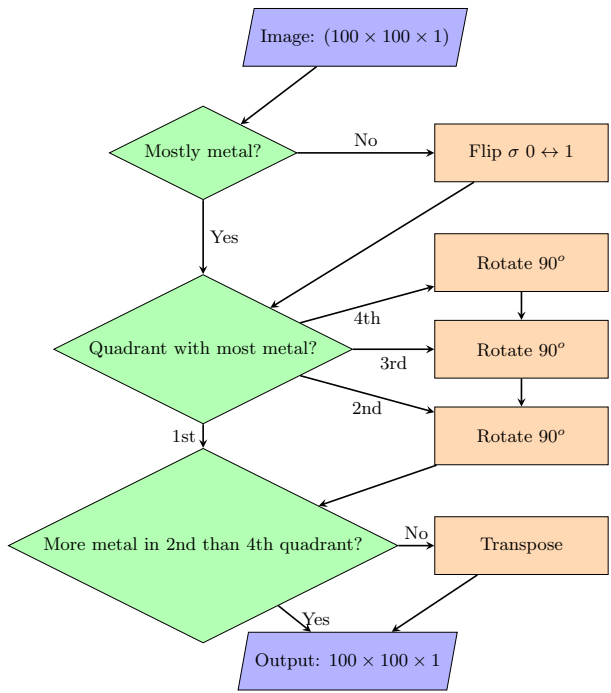

The parameters from Table \ref{tab:sim_params} are used to generate 8,000 images for each model near its transition, with the exception that
percolation  away from the 2D and 3D critical percolation strengths accounts for 16,000 images, for a total of  64,000 training images
of synthetic data.  We describe below the three major components
of our Deep Learning model:  (A)  data preparation via symmetry reduction; (B) a convolutional neural net with multiple layers; and (C) our method for judging the confidence of the classifier.

\subsection{Data Preparation:  Symmetry Reduction Method}
\label{ssxn:Symmetry_Operations}

The entire phase space associated with typical configurations
generated by the models above satisfies certain symmetries.
For example, the clean Ising  model (Eqn.~\ref{eqn:Ising_model}) 
satisfies the $Z_2$ symmetry ${\sigma_i} \rightarrow -{\sigma_i}$.
Similarly, the RFIM (Eqn.~\ref{eqn:Ising_model}) is symmetric under the
simultaneous operations
${\sigma_i} \rightarrow -{\sigma_i}$ with ${h_i} \rightarrow -{h_i}$.
Likewise, the percolation model is symmetric
under the simultaneous operations ${\sigma_i} \rightarrow -{\sigma_i}$ with
$p \rightarrow 1-p$.  
In addition, for the square \textcolor{black}{domain configurations}
we use as training data,
the statistical weight of typical configurations in phase space
is symmetric under all of the operations of the dihedral group of the square,
$D_4$.  
Such symmetries are often employed in ML via
a technique called data augmentation, in which all of the distinct symmetry 
operations are applied to specific configurations, in order
to generate more configurations, and thereby 
augment the training data.  
When a neural network is trained under this kind of augmented data set,
the resulting trained neural network respects all of the
symmetries of the underlying models which produced the training data,
rather than suffering from accidental asymmetries which mimic
the random nature by which the training data are produced.
The number of distinct symmetry operations available in our case
is that of $Z_2 \otimes D_4$, or $2 \times 8 = 16$.  
For the 
\textcolor{black}{square-shaped images of domain patterns that}
we generate, using this method of
data augmentation would increase the training set by a factor of 16.  

Rather than employ data augmentation, we introduce a new method:
{\em symmetry reduction}.
We prepare the data by reducing the symmetry of each configuration 
as much as possible
before feeding it into the neural network.  
This symmetry reduction is as effective as the data augmentation method,
but significantly reduces the time needed to train the neural network.
\textcolor{black}{In order for symmetry reduction to be effective, it is
essential that {\em all} data go through the symmetry reduction before being fed into the CNN (including training, validation, and any subsequent real-world data fed into the classifier).  
}

Let us turn our attention to the 
$Z_2 \otimes D_4$  symmetry operations in effect.
\textcolor{black}{
Our models (Eqns.~\ref{eqn:Ising_model} and \ref{eqn:RFIM}, including the non-interacting percolation limit where $J \rightarrow 0$) map metal and insulator domains to Ising spins $\sigma = +1$ for metal, and $\sigma = -1$ for insulator.  
}
There are  $2 \times 8 = 16$ symmetry operations that can be applied to 
these \textcolor{black}{spin configurations}
while preserving the weights of the
typical configurations in phase space.  We perform the 
following symmetry operations to each configuration in order to prepare the data:
\begin{enumerate}
    \item Ising $Z_2$ symmetry ${\sigma_i} \rightarrow -{\sigma_i}$:
    If a domain configuration has majority spin down, we flip all spins to make it majority spin up. 
    \item Rotations by $0,~\pi/2,~\pi,~3\pi/2$:  The configuration is rotated such that of the four quadrants, {quadrant~I} has the most spins up. 
    \item Transpose (reflection about the xy diagonal): If quadrant IV has more spins up than quadrant II, we transpose the configuration to ensure that quadrant II has more spins up than quadrant IV.  
\end{enumerate}
All the above operations are performed in the given order and the logic is summarized in 
Fig.~\ref{fig:symmetry_CNN}.\footnote{There are some configuration which cannot be mapped to a unique configuration using the  symmetry reduction operations described in the text.
For example, if exactly half the domains are metal, the first step cannot reduce it down to a unique configuration. 
But the average frequency with which that occurs is $\frac{^{N}C_{N/2}}{2^N}\approx1.592 \times 10^{3008}/1.995 \times 10^{3010}\approx 0.8\%$, 
where $N = 100\times 100$. }


\subsection{Convolutional Neural Net Architecture}
\label{ssxn:Convolutional_Neural_Net_Architecture}

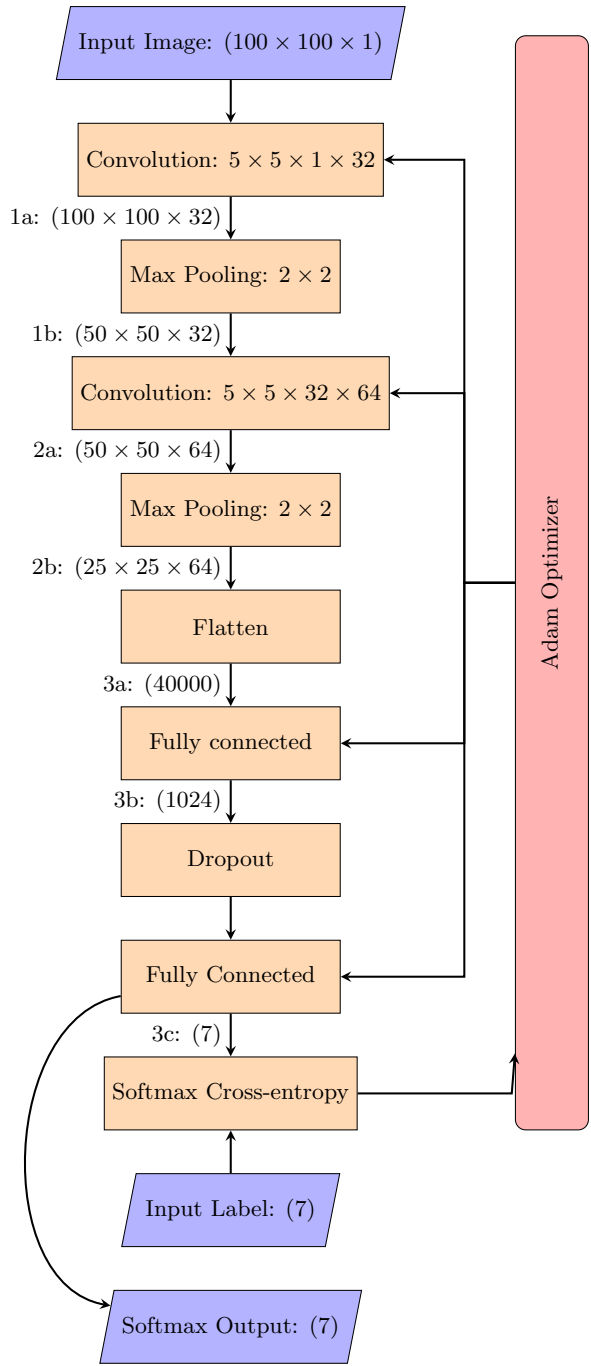
\begin{figure}
    \centering
    \resizebox{.45\textwidth}{!}{%
    \begin{tikzpicture}[node distance=1.6cm]
        \node (in1) [io] {Input Image: $(100\times100\times1)$};
        \node (pro1a) [process, below of=in1] {Convolution: $5 \times 5 \times 1 \times 32$} ;
        \node (pro1b) [process, below of=pro1a] {Max Pooling: $2 \times 2$} (pro1b);
        \node (pro2a) [process, below of=pro1b] {Convolution: $5 \times 5 \times 32 \times 64$} (pro2a);
        \node (pro2b) [process, below of=pro2a] {Max Pooling: $2 \times 2$} (pro2b);
        \node (pro3a) [process, below of=pro2b] {Flatten} (pro3a);
        \node (pro3b) [process, below of=pro3a] {Fully connected};
        \node (pro3c0) [process, below of=pro3b] {Dropout};
        \node (pro3c) [process, below of=pro3c0] {Fully Connected};
        \node (pro3d) [process, below of=pro3c] {Softmax Cross-entropy};
        \node (in2) [io, below of=pro3d] {Input Label: $(7)$};
        \node (out1) [io, below of=in2] {\textcolor{black}{Softmax} Output: $(7)$};
        \node (opt1) [startstop, right of=pro2b, rotate=90, xshift=-1cm, yshift=-2.8cm, minimum width=15cm] {Adam Optimizer};
        \draw [arrow] (in1) -- (pro1a);
        \draw [arrow] (pro1a) -- node[anchor=east] { 1a: $(100\times100\times32)$} (pro1b);
        \draw [arrow] (pro1b) -- node[anchor=east] { 1b: $(50\times50\times32)$} (pro2a);
        \draw [arrow] (pro2a) -- node[anchor=east] { 2a: $(50\times50\times64)$} (pro2b);
        \draw [arrow] (pro2b) -- node[anchor=east] { 2b: $(25\times25\times64)$} (pro3a);
        \draw [arrow] (pro3a) -- node[anchor=east] { 3a: $(40000)$} (pro3b);
        \draw [arrow] (pro3b) -- node[anchor=east] { 3b: $(1024)$} (pro3c0);
        \draw [arrow] (pro3c0) -- node[anchor=east] { } (pro3c);
        \draw [arrow] (pro3c) -- node[anchor=east] { 3c: $(7)$} (pro3d);
        \draw [arrow] (in2) -- (pro3d);
        \draw [arrow] (pro3c) to [out=190, in=170]  (out1);
        \draw [arrow] (opt1) - ++(-1.2,0) |- (pro1a);
        \draw [arrow] (opt1) - ++(-1.2,0) |- (pro2a);
        \draw [arrow] (opt1) - ++(-1.2,0) |- (pro3b);
        \draw [arrow] (opt1) - ++(-1.2,0) |- (pro3c);
        \draw [arrow] (pro3d) - ++(3,0) -- ++(3.85,0) -- (opt1);
    \end{tikzpicture}
    }
    \caption{%
    \label{fig:CNN_flow}
        Convolutional Neural Network. The input image here is reduced by the symmetry operations given in Fig.~\ref{fig:symmetry_CNN}. 
        The multi-dimensional output 2b is flattened into a one-dimensional array (3a)  before it is fed into the fully connected layer.
        We use the Adam (Adaptive moment estimation) optimization algorithm to train the network.~\cite{kingma2014adam} The output label is determined using softmax activation on the output layer.
    }
\end{figure}

\begin{figure}
    \begin{center}
        \includegraphics[width=.45\textwidth]{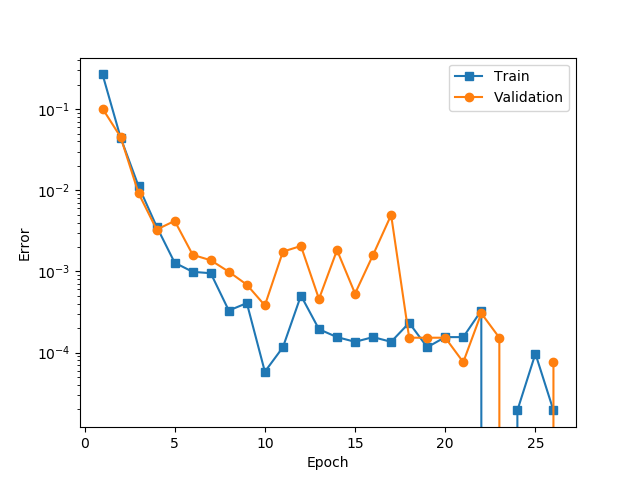}%
        \caption{%
        \label{fig:error_CNN}
            Error in the training and validation set vs. the number of epochs.
            Epochs correspond to the number of times the training set went through a training process.  
            To prevent overfitting we chose epoch=4 for testing with experimental images. Training/Validation accuracy = 99.64\%/99.67\%
        }
    \end{center}
\end{figure}

\begin{figure*}
    \begin{center}
        \includegraphics[width=.8\textwidth]{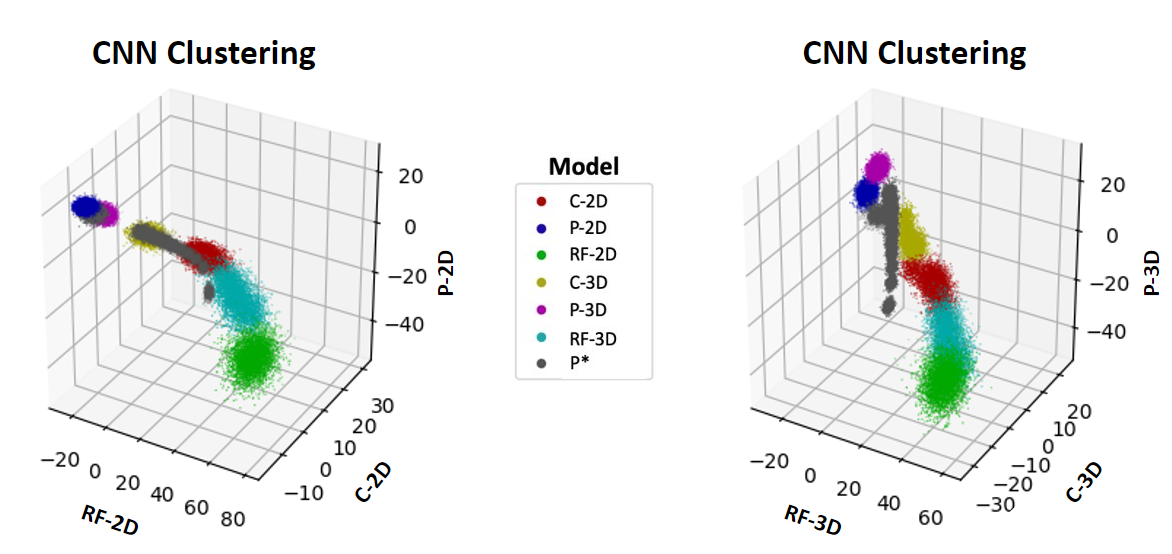} 
        \caption{%
        \label{fig:dist_out}%
            Distribution of 
            \textcolor{black}{values of the output nodes for
            }
            each class in the last fully connected layer, for all of the training sets.
             These clusters inhabit a $7$-dimensional ``model'' space.  The two representations in the figure are projections of the same $7$-dimensional information
             onto two different $3$-dimensional subspaces.
        }
    \end{center}
\end{figure*}


We pass to the neural network 
single channel binary images ({\em i.e.} strictly black and white, the same image space as QR codes) of size $100 \times 100$.
The architecture of the convolutional neural network (CNN) is as follows (Fig.~\ref{fig:CNN_flow}):
We use two sets of convolutional layers interleaved with max pooling layers.  
 The first convolutional layer applies a suite of 32 filters of size $5\times 5$ to the image, resulting in an image with 32 channels.
(By way of comparison, an RGB image has 3 channels, so that each pixel is described by 3 numbers.)
These 32 filters have a total of $32 \times 5 \times 5$ parameters to be trained.
The subsequent max pooling layer groups successive sets of $2 \times 2$ pixels,
keeping only the largest value in each channel, thus reducing the image size to $50 \times 50$.  
The next convolutional layer applies a suite of 64 filters of size $5 \times 5 \times 32$
to the 32-channel image which was passed from the previous max pooling layer.
These 64 filters have a total of $64 \times 5 \times 5$ parameters to be trained.  
This is followed by another max pooling layer, reducing the image size to 
$25 \times 25$ pixels, now with 64 channels.

All of this is 
followed by a fully connected layer, followed by dropout of 
\textcolor{black}{50 \%} 
of the connections, followed by a final fully connected layer, 
resulting in seven-dimensional output for classification
(C-2D, C-3D, RF-2D, RF-3D, P-2D, P-3D, and P*).
We use a softmax activation in the final output layer, which results in  single label classification.  If there are $n$ output classes with numbers $v_i$, the softmax activation function is defined as:
\begin{equation}
Y_i^{softmax} = \exp{(v_i)}/\sum_{j=1}^{n}\exp{(v_j)}
\label{eqn:softmax}
\end{equation}
where $Y_i^{softmax}$ is the output likelihood estimate.

After the symmetry reduction, 80$\%$ of the configurations are used
for training;  the remaining 20$\%$ are used for validation. 
The training set is used to train the network whereas the validation set is used 
to predict the expected error upon generalization beyond the training set.
Figure \ref{fig:error_CNN} shows how the errors evolve with training epoch.
The epoch at which the classification errors in the validation phase deviate from the errors in the training phase roughly marks the onset of overfitting. 
Figure \ref{fig:error_CNN} shows that the classification errors are less than 0.5\% at this point.


\subsection{Figure of Merit of Classifications}
\label{ssxn:Rejection_Criteria}

Our goal is to use our ML model developed under supervised learning conditions to distinguish among hypotheses about datasets from real experiments.  
But before applying our trained ML model to images from experiment,
it is important to understand that a trained classifier is only
as good as its training set.
Thus far, we have generated 
“simulated data” from various theoretical Hamiltonians, and we have trained an ML algorithm as to which sets of simulated data came from which underlying Hamiltonian.  
A major challenge in going from simulated data to real-world data is how  to control for hypotheses that were not originally envisioned.  For example, if an ML classifier has been trained to recognize the difference between cats and dogs, what answer will it give when shown a banana?  A simple classifier will give a classification from its training set, but 
ideally, the answer should not be ``cat'' or ``dog,'' but rather, ``neither.''
Likewise, if our ML classifier is shown experimental data from a system whose
underlying Hamiltonian is sufficiently different that none of our Hamiltonians
used in the training process are a good description of the physical system,
a simple classifier will still return {\em some} classification.
Therefore, it is necessary to devise a method for flagging potentially
dubious classifications.  
One method is adversarial training, {\em i.e.} to train the CNN on images that are not in the set of Hamiltonians comprising the hypothesis.  Once again, this is limited by human imagination.  For example, how will one know when this process is sufficiently completed, and how can one control for unforeseen image types arising in experiment?  It is better to design a neutral method for flagging suspicious classifications, one that is not limited by the adversarial training set.  

Therefore, we seek to devise a completely different method for identifying potentially dubious classifications.  In order to do this, we turn our attention to the distribution of values observed right after the last fully connected layer in Fig.~\ref{fig:CNN_flow}.
Fig.~\ref{fig:dist_out} shows what the distribution of values looks like at this step, over the entire training set.  
Since this distribution is well clustered for the seven models of interest, a prediction point lying far from its corresponding cluster should 
be scrutinized rather than blindly accepted.

For each class, the distribution at the end of the last fully connected layer (see Fig.~\ref{fig:dist_out})
is generated from the training examples.
We form the $7$-dimensional standard deviation vector of these clusters about their centers of mass.  
We subsequently {\em flag as suspicious} any output in this layer that is 
a distance in this space of more than one standard deviation vector from all points in the cluster.  
Setting the cutoff at smaller distances rejects too many correct predictions in the validation set.
A generalization of this method would be to use any or all of the intermediate layers for detecting such an anomaly in the input data, 
see Ref.~\cite{sivamani2020nonintrusive}.



\section{Application to Experimental Images on VO$_2$}
\label{sxn:Application_to_Experimental_Images}

\subsection{Testing the CNN on SNIM images of a thin film of VO$_2$}
\label{ssxn:Classifying_SNIM_images_of_VO_2_film}

We next turn our attention to testing
the trained CNN on an experimentally derived dataset
for which the Hamiltonian underlying the 
experimentally observed pattern formation is already known, 
before applying the CNN to a new experimental dataset
for which the answer is not previously known.
In this section, we consider experimental data 
taken via SNIM on a thin film of VO$_2$.
VO$_2$ undergoes a metal-insulator transition just above room temperature,
in which the resistivity changes by over five orders of magnitude.\cite{Morin1959}
Rather than transitioning all at once,
\textcolor{black}{we previously showed}
that there is a finite regime of phase coexistence in which the
metal and insulator puddles show significant pattern formation.\cite{qazilbash-science}
\textcolor{black}{In fact, the spatial correlations reveal structure
on all length scales  measured via SNIM, 
from the pixel size (20nm) all the way out to the field of view (4 $\mu m$).\cite{shuo-prl} }
The physics driving the pattern formation in this sample is already known via
the cluster analysis techniques we recently developed.\cite{Phillabaum:2012cv,superstripes-2014,C-3Dx}
By applying these techniques to analyze the metal and insulator puddles
in this thin film of VO$_2$, we showed that the multiscale domains
are of a fractal nature, with 
quantitative geometric characteristics including avalanche statistics
matching those of the RF-2D.\cite{shuo-prl}

\begin{figure*}
    \begin{center}
        \includegraphics[width=\textwidth]{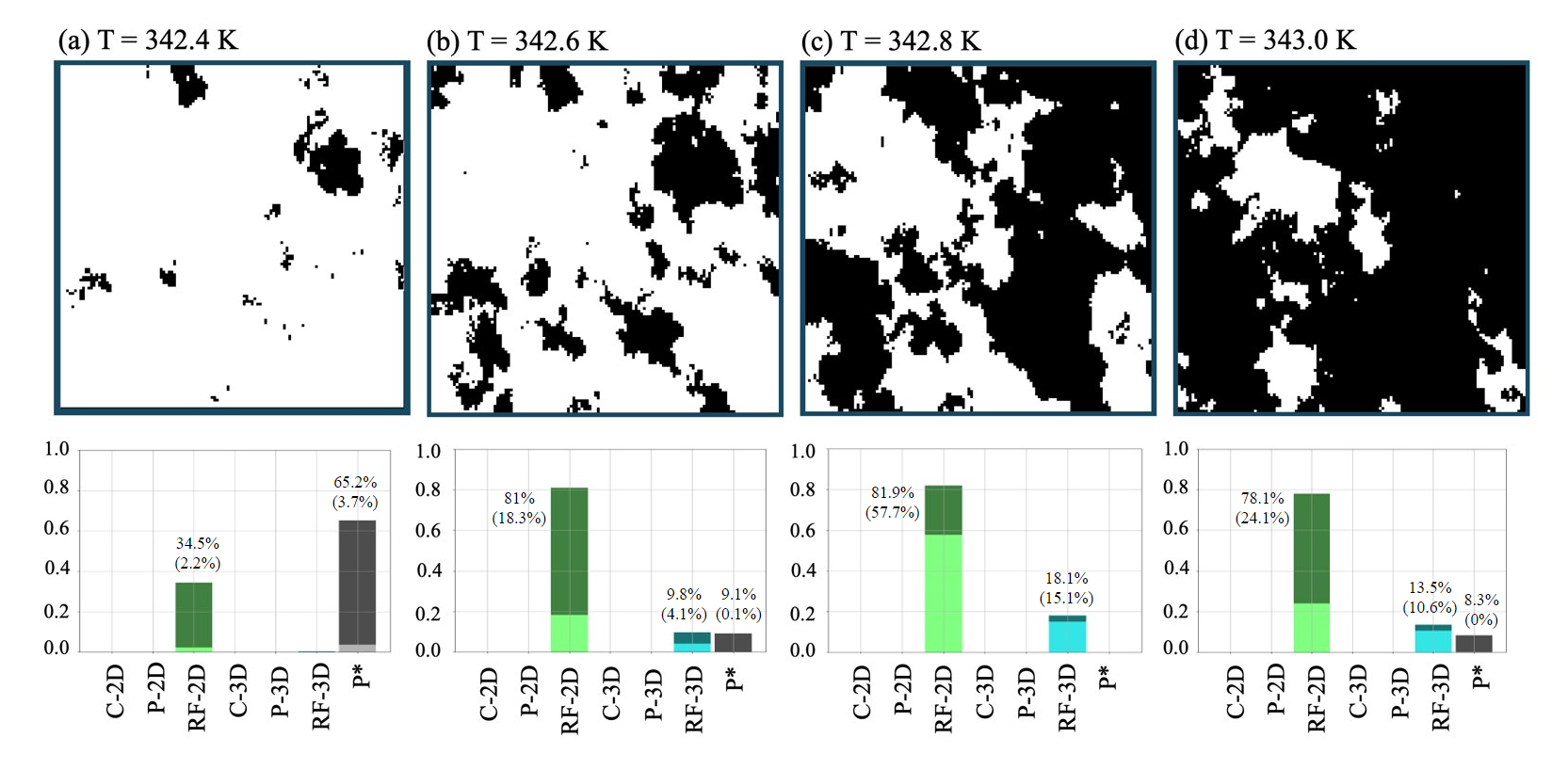}
    \caption{%
    \label{fig:MQ_soft_b}
        Classification results of our deep learning model applied to SNIM images on a thin film of VO$_2$ as described in the text.  
        The top row shows the 
        thresholded data as described in the text.  
        The field of view is $4 \mu m \times 4 \mu m$.
        White patches are insulating; black patches are metallic.  
        The total percentage of classifications for a particular
        model are reported in the bar charts of panels (a)-(d).
        Classification percentages that fall within $1 \sigma$ of a cluster in the training set are indicated in parentheses.  
        Classifications that fall more than $1 \sigma$
        away from the edge of the corresponding cluster
        in the training set are colored darker in the bar chart.
    }
    \end{center}
\end{figure*}

Fig.~\ref{fig:MQ_soft_b} 
shows the application of the CNN to experimental data on a thin film 
of VO$_2$ as it undergoes the metal-insulator transition.
The data were obtained using SNIM,
and first reported in Ref.~\onlinecite{qazilbash-science}.
SNIM  measurements return an intensity $a$
as a continuous variable at each pixel, resulting in single channel
images.   These SNIM images are of size 256 px $\times$ 256 px.
The SNIM  images are converted to black pixels and white pixels
by assigning SNIM values of $a < 2.5$, which are insulating, to be white, and SNIM values of
and $a > 2.5$, which are metallic, to be black,
as discussed in Ref.~\cite{shuo-prl}.
These 
thresholded images are shown in 
the top row of Fig.~\ref{fig:MQ_soft_b}.  
Ref.~\cite{shuo-prl} showed that the geometric characteristics
of the pattern formation are insensitive to changes in 
the threshold $a$ within about 15\% of this threshold value.

The CNN takes in the images 100px $\times$ 100px at a time, 
in each instance returning a classification indicating which
Hamiltonian likely produced the pattern formation.  
In order to make full use of the spatial structure in the image,
we use a sliding window of size 100px $\times$ 100px, resulting in  
\textcolor{black}{$ (256-100+1) \times (256-100+1) = 157 \times 157 $}  
classifications for each image. 

\begin{figure*}
    \begin{center}
        \includegraphics[width=0.7\textwidth]{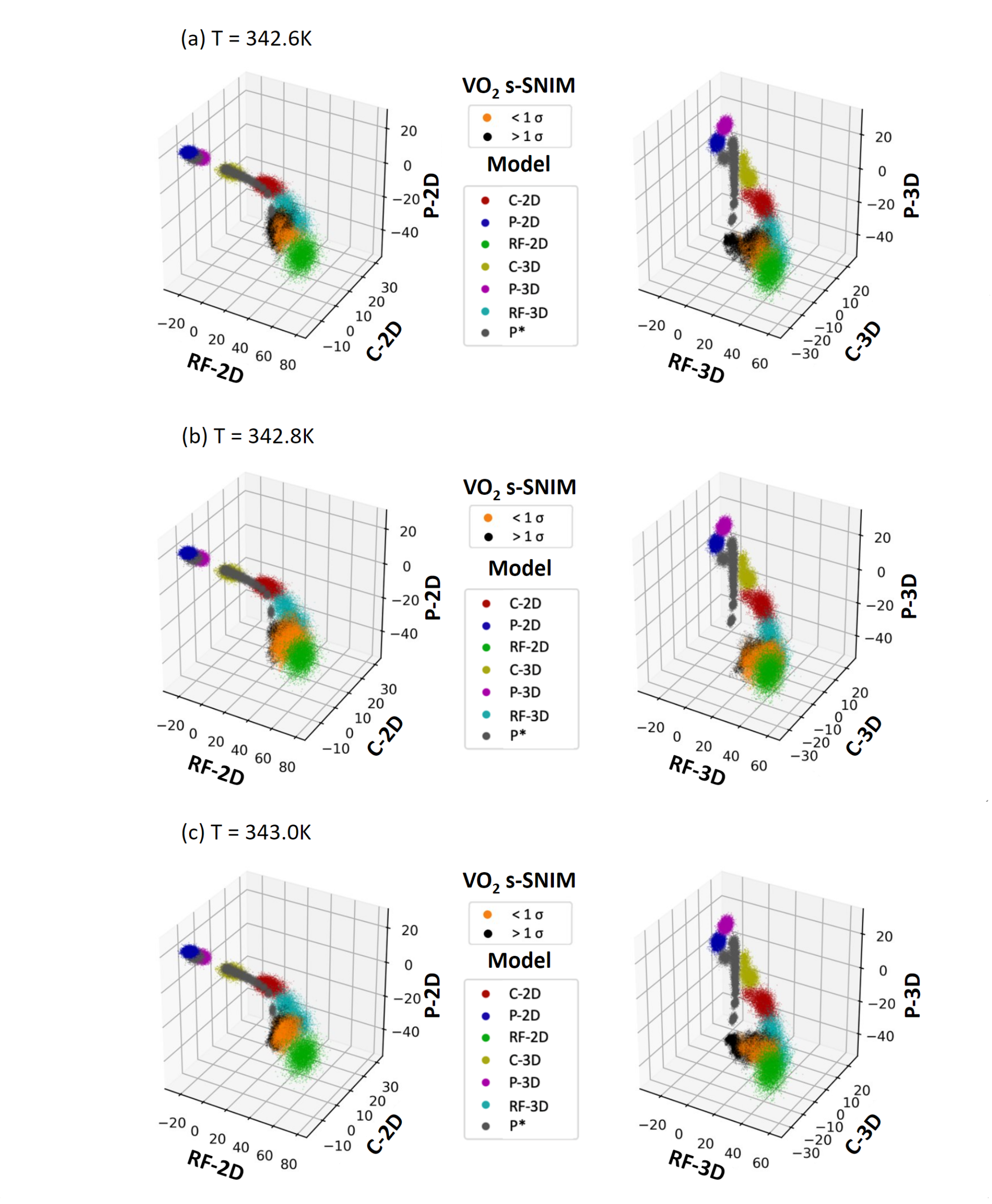}
    \caption{%
    \label{fig:MQ_soft_d}
        Distribution of 
        \textcolor{black}{values of the output nodes for
        }
        each class in the last fully connected layer, for the VO$_2$ SNIM data, superimposed on the distribution for the training sets shown in Fig.~\ref{fig:dist_out}.
        Results for the VO$_2$ data that are within one $7$-dimensional standard deviation of a training set are indicated by orange  dots.
       Results for the VO$_2$ data that are farther away are indicated by black dots.   
    }
    \end{center}
\end{figure*}

In Fig.~\ref{fig:MQ_soft_d}, 
we show the distribution of values 
in the last fully connected layer of the CNN,
over the set of sliding windows.
The small circles correspond to the training set,
and are the same as those shown in Fig.~\ref{fig:dist_out}.
The large circles are the result of the CNN applied to 
the experimentally derived SNIM images.  
If the large circles lie within one standard deviation
of any point in the corresponding training cluster, 
they are colored black.
If the large circles are not within one standard deviation
of at least one point within the corresponding validation cluster,
they are labeled gray, 
as described 
in Sec.~\ref{ssxn:Rejection_Criteria}.


Fig.~\ref{fig:MQ_soft_b} shows the final results of the classifier
applied to the SNIM data.  
Below each SNIM image (top row), the bar chart indicates the
percentage of sliding windows which give a particular classification.  
Bright bars and numbers in parentheses correspond to classifications 
that are within one $7$-dimensional standard deviation of the training sets.
The darker part of the bar, and the numbers not in parentheses refer to the
the total percentage of sliding windows which give the corresponding
classification.  Thus the overall result of the ML classifier 
on an experimentally derived dataset is that it agrees with the 
classification from cluster techniques, with at least 83$\%$ confidence.

Notice that in Fig.~\ref{fig:MQ_soft_d}, 
the distribution of 
\textcolor{black}{values of the output nodes
}
in the last fully connected layer for the CNN applied to the SNIM data is always 
close to the RF-2D model. Moreover, the entire set of points moves toward
the training set distribution and then away from it, as a function
of temperature.  The temperature of closest approach is T = 342.8K.  
This same phenomenon is borne out in the bar charts
of Fig.~\ref{fig:MQ_soft_b}, where height of the bright green bar also
peaks at T = 342.8K.  
This is highly reminiscent of critical behavior,
which grows in strength as the system approaches criticality, and
diminishes as the system moves away from criticality.  
We propose that the distance of the center of mass of the 
SNIM cluster from the training clusters 
can be used as a measure of proximity to the critical point.
Further study is needed to test this idea.   


\begin{figure*}
    \begin{center}
    \includegraphics[width=\textwidth,trim={0.0cm 0.0cm 0.2cm 0.0cm},clip]{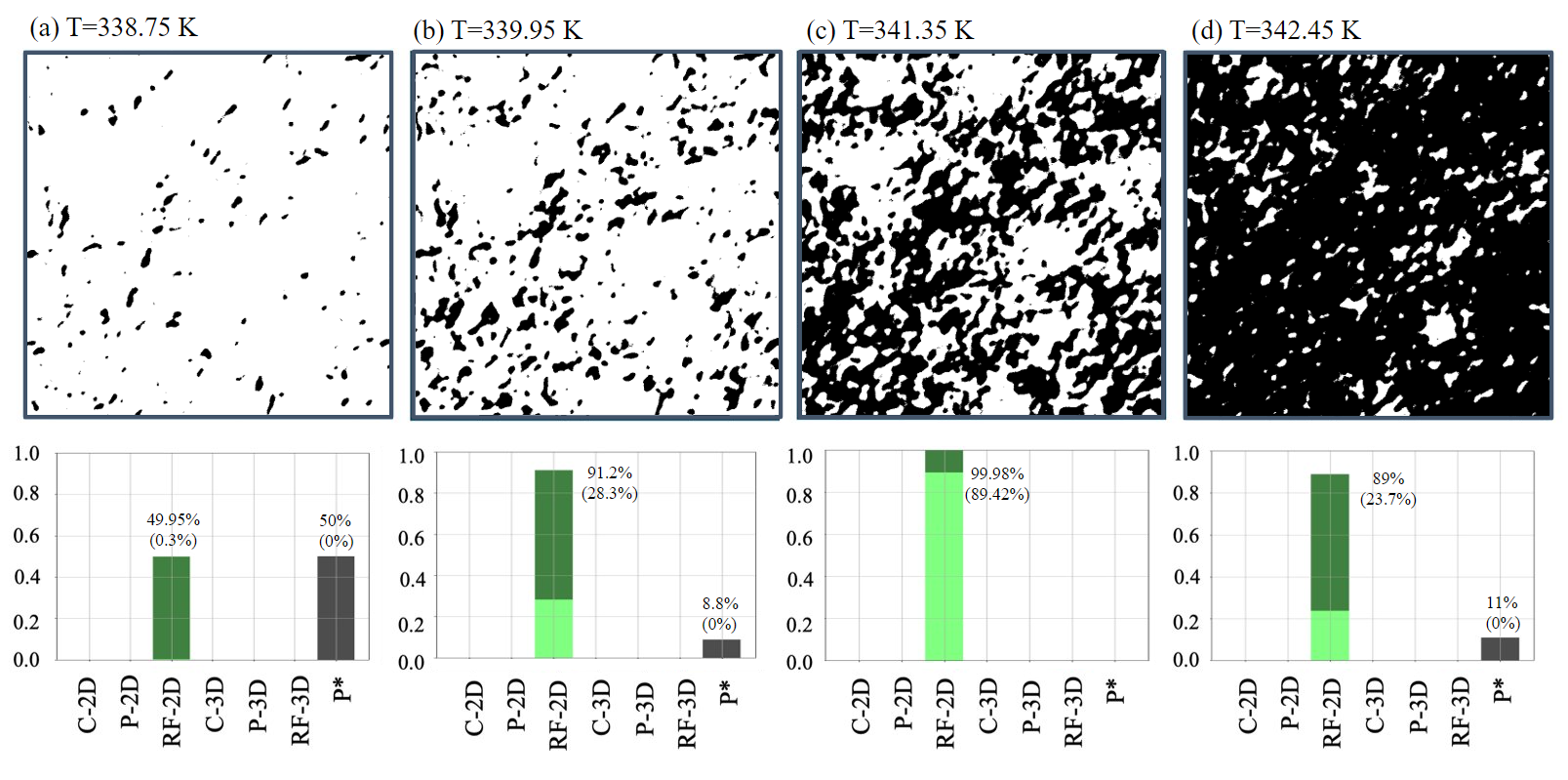}
    \caption{%
    Classification results of our deep learning model applied to new 28$\mu m \times 28\mu m$ optical microscopy images of a VO$_2$ thin film as described in the text.
     White patches are insulating; black patches are metallic.  
        The total percentage of classifications for a particular
        model are reported in the bar charts of panels (a)-(e).
        Classification percentages that fall within $1 \sigma$ of a cluster in the training set are indicated in parentheses.  
        Classifications that fall more than $1 \sigma$
        away from the edge of the corresponding cluster
        in the training set are colored darker in the bar chart. All CNN predictions from optical data during a full temperature ramp up are presented in SI Fig.~\ref{fig:film_full_Melissa}.
    }
        \label{fig:film_Melissa}
    \end{center}
\end{figure*}


\subsection{Applying the CNN to New Optical Microscope Images of a VO$_2$ film}
\label{ssxn:Classifying_Optical_Microscope_images_of_VO_2_film}

The top panels in Fig.~\ref{fig:film_Melissa} 
show metal and insulator domains in a thin film of VO$_2$
made at UCSD, taken using a home-built 
optical microscopy system capable of remaining in focus while temperature is cycled through the full metal-insulator transition. \textcolor{black}{(See SI and Ref.~\cite{instrument_optics} for full details
of the sample preparation and experimental setup.)} The optical data are taken 
at a series of temperatures going through the metal-insulator transition.  
\textcolor{black}{
The physical dimensions of the square image sizes in  Fig.~\ref{fig:film_Melissa} are all 28$\mu m \times 28\mu m$,
and the pixel size is 50nm.\footnote{The far field optical resolution is 300-500nm.}  Both the FOV and the pixel size
are larger than those of the SNIM images
in Fig.~\ref{fig:MQ_soft_b}.  
}

We apply the same sliding window technique 
as with the SNIM data
to analyze pieces of each image, $100$px $\times$ $100$px at
a time,  in each instance returning a classification indicating which Hamiltonian likely produced the pattern formation.  
Because the optical images in 
Fig.~\ref{fig:film_Melissa}
are
$ 760 \times 760 $ pixels,
this results in $ (760-100+1) \times (760-100+1) = 661^2$ classifications for each image.

\begin{figure*}
    \begin{center}
        \includegraphics[width=0.7\textwidth]{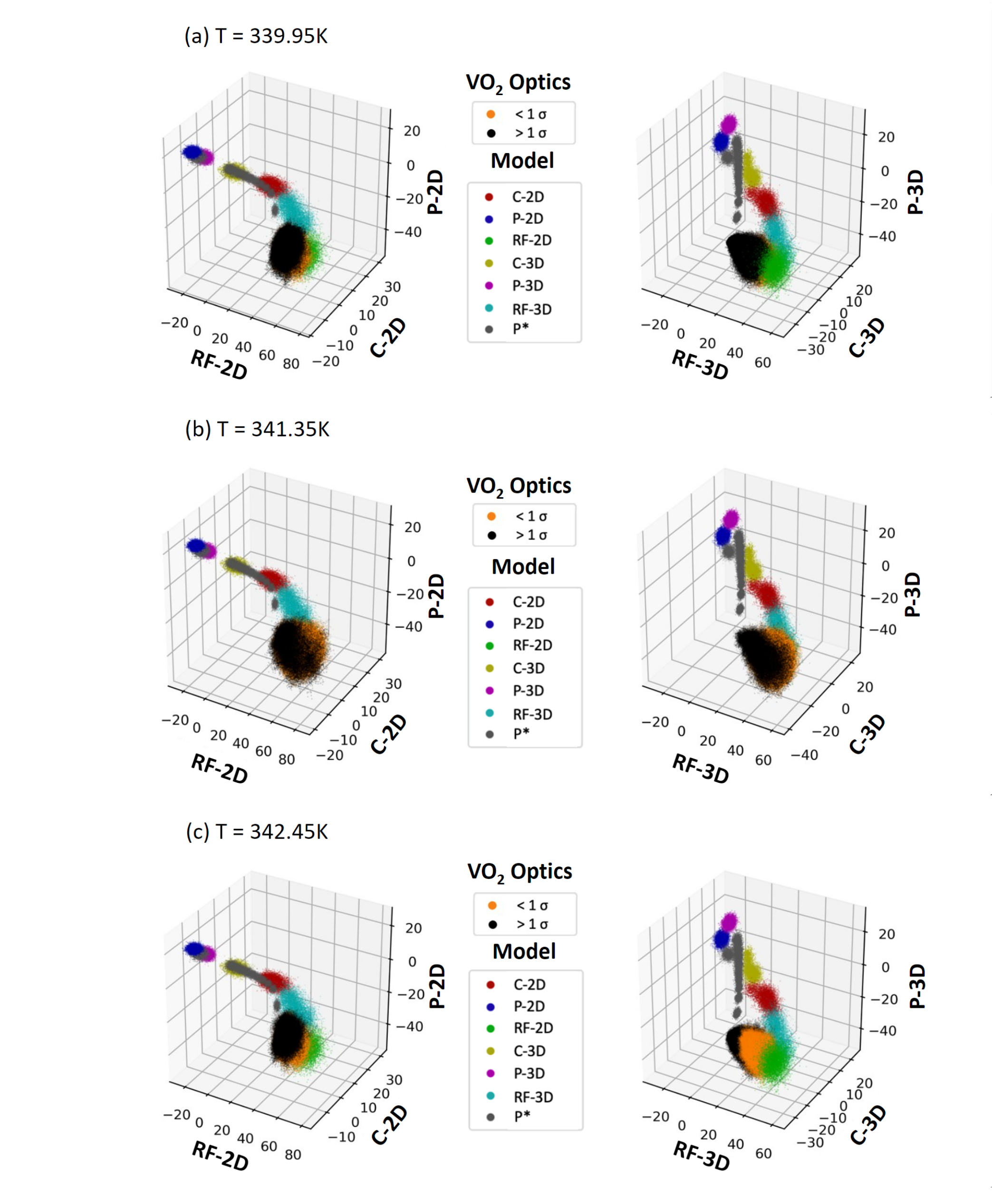}
    \caption{%
    \label{fig:MQ_soft_d_optics_square}
        Distribution of relative weights of each class in the last fully connected layer, for the VO$_2$ optical data (square sample presented in the main text Fig.~\ref{fig:film_Melissa}), superimposed on the distribution for the training sets shown in Fig.~\ref{fig:dist_out}.
        Results for the VO$_2$ data that are within one $7$-dimensional standard deviation of a training set are indicated by orange  dots.
       Results for the VO$_2$ data that are farther away are indicated by black dots.   
    }
    \end{center}
\end{figure*}

The bottom panels in
Fig.~\ref{fig:film_Melissa}
show the final results of the CNN classifier
applied to the optical microscope data.  
Below each optical microscope image, the bar
chart indicates the percentage of sliding windows which
give a particular classification.  
In this case the images from temperatures T = 339K to
T = 343K are each identified as RF-2D with a maximum greater than 89$\%$ confidence.
In Figure~\ref{fig:MQ_soft_d_optics_square} 
we show the distribution of values in the last fully connected layer of the CNN,
over the set of sliding windows.
The small circles correspond to the training set,
and are the same as those shown in Fig.~\ref{fig:dist_out}.
We discuss the implications of this identification in 
Sec.~\ref{sxn:Discussion}.

\begin{figure*}
    \begin{center}
    \includegraphics[width=.49\textwidth]{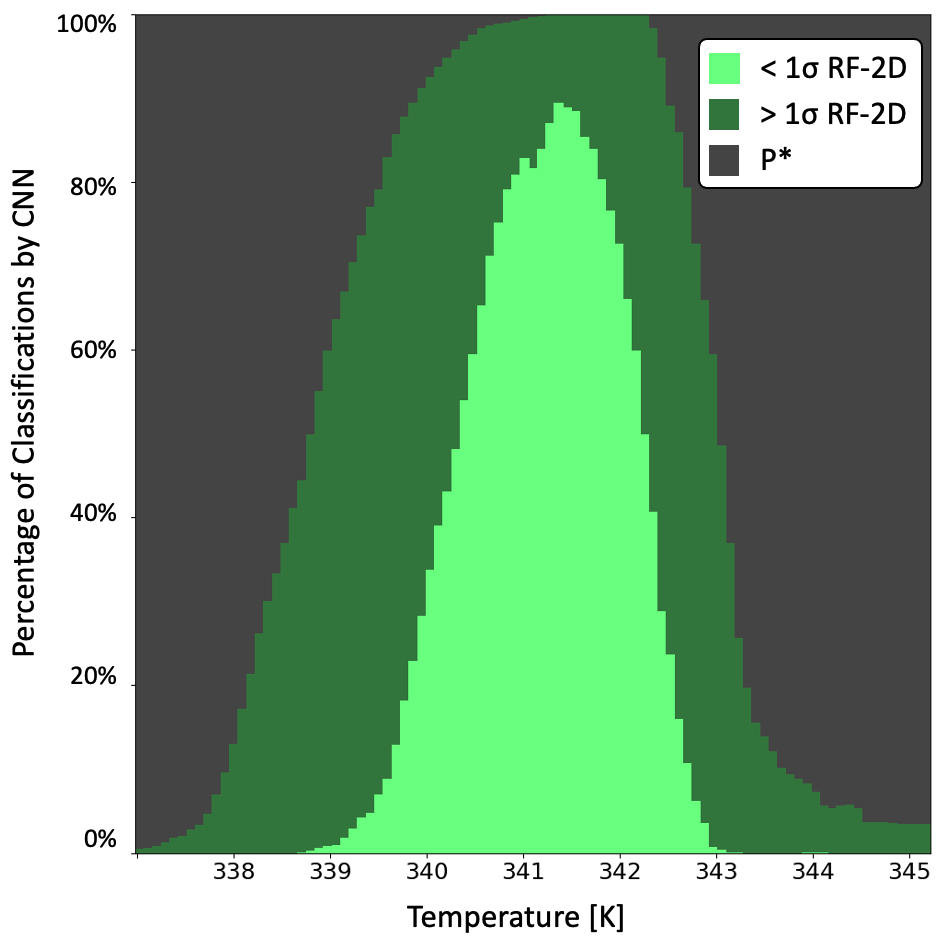}%
    \caption{%
        All CNN predictions from optical data during a temperature ramp up of data presented in Fig.~\ref{fig:film_Melissa}.
        Darker colors denote classifications that are more than 1 standard deviation from the identified training set.
    \label{fig:film_full_Melissa}}
    \end{center}
\end{figure*}

From a theoretical point of view,
we don't expect {\em every} image acquired from the experiments 
to have significant pattern formation. 
For example, once the image saturates to metal or insulator, there is no pattern formation left, and consequently there is  much less information available in these datasets about the underlying model.  Rather, we expect the images to display criticality which reaches peak prominence at a particular temperature.  
The typical method to discern proximity to criticality is through correlation lengths.
The correlation length is expected to blow up as a power law, $\xi \propto 1/|T-T_c|^{\nu}$ in the vicinity of the critical temperature.  
However, the maximum correlation length our CNN can discern is cut off 
by the maximum FOV that the CNN is fed from the experimental data.
Furthermore, the CNN analysis does not return a length scale.
Instead, we observe once again the interesting behavior that the 
proximity of the experimentally derived data's cluster of 
\textcolor{black}{output values}
in the last fully connected layer approaches and then retreats
from the cluster of 
\textcolor{black}{output values}
in the training sets as a function of temperature,
as evidenced by the non-monotonic behavior of the height of the bright
green bars with temperature in Fig.~\ref{fig:film_full_Melissa}.
This is in line with our previous conjecture that the average distance
of the cluster from that of the training set can be
used as a measure of proximity to criticality.  

\section{Discussion}
\label{sxn:Discussion}

For both experimental datasets, whether from SNIM
or from optical microscopy, the deep learning CNN
determined that the intricate pattern of metal and 
insulator patches was being set by the physics of the
RF-2D.  
\textcolor{black}{For the SNIM data, this matches our prior 
identification using cluster methods.\cite{shuo-prl}
For the microscope data, 
}
it was
already known prior to application of the CNN that the
physics driving the pattern formation should be arising
from a 2D Hamiltonian. 
\textcolor{black}{
This is because the thickness of the film ($\approx 300nm$) 
is comparable to the lateral resolution of the instrument.
Consequently, the spatial correlations being measured are
firmly in the two-dimensional limit. 
}
However, the fact that the CNN returned a two-dimensional model
\textcolor{black}{and not a three-dimensional model}
gives us further confidence in the CNN method.

The identification of the Hamiltonian as RF-2D means that
a combination of material disorder and interactions between spatially proximate
regions of the sample drives the pattern formation.  
\textcolor{black}{The fact that interactions must be present rules out a Preisach
model\cite{preisach} 
of independent hysteretic switchers, as we previously argued based on first-order reversal curve measurements\cite{schuller-FORC}
and a cluster analysis of the critical exponents 
during the transition.\cite{shuo-prl}
}
\textcolor{black}{
The multiscale nature of the pattern formation is driven by
proximity to criticality, which can happen even in a 1st order
phase transition, near a critical endpoint.\cite{shuo-prl}
\footnote{Interestingly, our identification of RF-2D means that in fact the 1st order line has gone away, since the critical disorder strength in 2D is 0.  However, hysteresis loops remain open because of the extreme critical slowing down of random field Ising models, which generate exponentially large timescales for equilibration.}
Consistent with proximity to criticality, we have previously shown
that there is significant slowing down of the relaxation time
near the phase transition.\cite{schuller-ultrafast,schuller-longtime}
}

Random field 
critical points exhibit extreme critical slowing down: 
because the barriers to equilibration grow as a power law as the system nears criticality, the characteristic relaxation time grows exponentially as the system
approaches criticality.\cite{dan-fisher-rfim}
Because of this extreme critical slowing down, the model is notorious for highly nonequilibrium behavior, including hysteresis, glassiness, coarsening, and aging.
In addition, the model has an anomalously large region of critical behavior:
a system that is 85$\%$ away from the critical point 
can still display
2 decades of scaling.\cite{perkovic}  This means that it is fairly
easy within this model to get into a regime that displays pattern formation
across multiple length scales, including fractal textures.  

\textcolor{black}{
With the model controlling this pattern formation now well established
from this study and from our previous work\cite{shuo-prl}, 
we can make the following statements about VO$_2$:
Increasing disorder is expected to broaden hysteresis curves,
and also decrease the slope of the hysteresis curve at its inflection point.\cite{dahmen-prl-1993}  
Indeed, these expectations are borne out in recent
ion irradiation studies of resistivity in VO$_2$.\cite{schuller-lesueur-disorder}
}
In addition, due to the pronounced
memory effects with exponentially long equilibration times, 
exact identification of material properties can be history-dependent,
leading to the appearance of non-repeatability.  
On the other hand, disorder can ultimately be exploited as another means of control\cite{dagotto-science,nat-comm-2011}.

The ML method is complementary to the aforementioned cluster techniques.  Whereas the cluster techniques require at least two
decades of scaling in the dataset, we have shown here and in 
Ref.~\cite{lukasz-ML}
that an ML classifier can make determinations on datasets 
with smaller FOV.
And while the cluster techniques are designed to extract information
from datasets in systems that are in the vicinity of a critical point,
we expect that the ML methods developed here can be useful farther away
from criticality, because they are able to make determinations on smaller FOV, {\em i.e.} they do not require that the system have the long correlation length
associated with proximity to criticality.  

In the same way that 
the critical exponents are  encoded in the shapes and statistics of the fractal electronic textures that arise near a critical point,\cite{Phillabaum:2012cv,superstripes-2014,shuo-vo2}
our ML study reveals that the universal features of the model itself are encoded in the spatial correlations of the textures, without needing the
intermediate step of identifying critical exponents.  Criticality presents itself even at the moderate ({\em i.e. not long}) length scales our CNN views, which is set by the size of the sliding window we employ on the datasets to be classified.  

The method is also potentially extendible to handle non-discrete order parameters, such as continuum models, which present a challenge for cluster methods.  For example, it may be possible
to use a similar framework to diagnose pattern formation that reveals
an underlying XY model or Heisenberg model.  
In addition, by using regression, we expect to be able to go beyond criticality to begin to determine the values of parameters in the Hamiltonian.   

We have developed this ML method first on critical systems, which have no characteristic length scale due to the power law structure, and therefore display spatial structure on every length scale within a correlation length.  
However, we expect this general scheme to also be broadly applicable to systems that have an emergent length scale, such as frustrated phase separation systems in general, such as block copolymers, the mixed phase of Type I superconductors, reaction-diffusion systems, and convection rolls.\cite{seul-andelman,kivelson-frustrated}

Dagotto\cite{dagotto-science} points out that quenched disorder plays an important role in many strongly correlated materials, and based on this he argues that for such materials,  ``it is not sufficient to consider phase diagrams involving only temperature and hole-doping x.  A  disorder strength axis should be incorporated into the phase diagram of these materials as  well.''
Models incorporating disorder predict that nonequilibrium behavior including glassiness (multiple nearby local energy minima) and hysteresis are prominent features when electronic phase separation occurs in the presence of quenched disorder.\cite{Phillabaum:2012cv,superstripes-2014,shuo-vo2,detecting-nematics,rfim-prl,loh-stm-noise,post-natphys,jiarui-simmons,dahmen-prb-1996,sethna-prl-1993}    
The methods we have employed here, which identify the terms in the Hamiltonian, when extended to include a regression analysis to identify the values of the parameters in those terms, 
have the potential to identify the disorder strength.
Mapping out this disorder strength axis in strongly correlated
phase diagrams has the potential to help disentangle some of the
ambiguities and apparent inconsistencies heretofore reported in the literature of these systems.\cite{pseudogap-timusk,pseudogap-armitage,kivelson-keimer,bozovic-levy}

Future work on this type of classifier will also benefit from:
(1) generalizing the CNN to handle input images of any size;
(2) developing a learning-based optimization for the rejection classifier;
and (3) handling grayscale images without the need to threshold them.

\section{Conclusion}
\label{sxn:Conclusion}


In conclusion, we have extended machine learning methods to 
be able to identify the Hamiltonian driving pattern formation
in complex electronic mater.  
We have shown the accuracy that can be achieved by using a convolutional neural net to classify synthetic data is better than 99\%, and about 83--89\% accurate on experimental data.
We introduce a symmetry reduction method, which significantly lowers the training time over data reduction without
reducing accuracy.
In addition, we introduce a distribution-based method for quantifying confidence of multilabel classifier predictions, without the problems associated with introducing adversarial training sets.
We also propose a new machine learning based criterion for diagnosing proximity to criticality.

We have also demonstrated that this framework can be successfuly applied to real experimental images, by using it to classify the Hamiltonian 
of SNIM data on a thin film of VO$_2$, for which the answer
was already known from a complementary theoretical method.  
Having thus vetted our ML model, we applied it to optical microscope
data on a different sample of VO$_2$.  In each case, we find that
the pattern formation of metal-insulator domains in thin films of
VO$_2$ is driven by proximity to a critical point of the two-dimensional
random field Ising model.  Further tests of this model include hysteresis protocols in the presence of a series of engineered disorder strengths.

\acknowledgments
We thank M.~J.~Carlson for technical assistance with image stabilization, and acknowledge helpful conversations with 
A.~El Gamal and K.~A.~Dahmen.
The work at ESPCI (M.A.B., L.A., and A.Z.) was supported by
Cofund AI4theSciences hosted by PSL Universit\'e,
through the the European Union’s Horizon 2020 Research and Innovation Programme under the Marie Skłodowska-Curie Grant No. 945304.
The work at UCSD (P.S. and I.K.S.) was supported by the AFOSR Award No. FA9550-20-1-0242. 
M.M.Q. acknowledges support from the National Science
Foundation (NSF) via Grant No. IIP-1827536.
D.N.B acknowledges support by the Center on Precision-Assembled Quantum Materials, funded through the US National Science Foundation (NSF) Materials Research Science and Engineering Centers (Award No. DMR-2011738).
S.B., F.S., and E.W.C. acknowledge  support from NSF Grant No. DMR-2006192,  NSF XSEDE Grant Nos. TG-DMR-180098 and DMR-190014, and the Research Corporation for Science Advancement Cottrell SEED Award.
F.S. acknowledges support from the COVID-19 Research Disruption Fund at Purdue
through the U.S. Department of Education HEERF III (ARP) Award No. P425F204928.
S.B. acknowledges support from a Bilsland Dissertation Fellowship. E.W.C. acknowledges support from a Fulbright Fellowship and from DOE BES Award No. DE-SC0022277.
L. B. acknowledges support from a Summer Undergraduate Research Fellowship at Purdue.
This research was supported in part through computational resources provided by Research Computing at Purdue, West Lafayette, Indiana.\cite{rcac-purdue}


%

\clearpage
\onecolumngrid
\setcounter{equation}{0}
\setcounter{figure}{0}
\setcounter{table}{0}
\setcounter{section}{0}
\setcounter{page}{1}
\renewcommand{\theequation}{S\arabic{equation}}
\renewcommand{\thefigure}{S\arabic{figure}}
\renewcommand{\thetable}{S\Roman{table}}
\renewcommand{\thesection}{S\arabic{section}}

\begin{center}
\textbf{\large Supplementary Material for\\
\bigskip
\mytitle}
\end{center}

\section{Experimental Methods}

\subsection{Sample Preparation}

The VO$_2$ film measured by optical microscopy was grown by reactive rf sputtering using a stoichiometric V$_2$O$_3$ target. The substrate temperature during the growth was 540$^o$C. The growth was done in 3.4 mTorr 92.3\% Ar 7.7\% O$_2$ atmosphere. The rf power was 100 W resulting in ~2.5 nm/min growth rate. After the growth, the sample was cooled down at 12$^o$C/min to room temperature while maintaining the Ar/O$_2$ atmosphere to preserve proper oxygen stoichiometry in the synthesized film. The film was patterned into 50 $\mu m \times $ 35 $\mu m$ patches by reactive ion etching in Cl$_2$/Ar atmosphere. Ti/Au electrodes defining 30$\mu m \times $35 $\mu m$  devices were made using standard lithography procedure and e-beam evaporation.

\subsection{Far Field Optical Microscopy}
Data for Fig.~\ref{fig:film_Melissa}  in the main text 
are taken 
using a home-built 
optical microscopy system capable of remaining in focus while temperature is cycled through the full metal-insulator transition.
The 300$nm$ thick film of VO$_2$, grown at UCSD, 
was deposited by rf magnetron sputtering 
on an r-cut sapphire substrate.
Gold electrodes separated by 30$\mu$m were deposited on top of the 35$\mu$m etched film \cite{instrument_optics}.
As temperature is ramped from 35$^o$C to 80$^o$ at 0.5$^o$C/min, the sample shows a clear insulator to metal transition around 68$^o$C, as evidenced by a change in its resistivity by four orders of magnitude.
The images in Figs.~\ref{fig:film_Melissa} and~\ref{fig:ESPCI-VO2_92} were taken using a home-built 
optical microscopy system capable of remaining in focus while temperature is cycled through the full metal-insulator transition.
Surface reflection images were taken in the visible range with a $\times150$ magnification dry Olympus objective lens with an optical aperture of 0.9 and a focal point of 1mm. The theoretical lateral resolution is estimated to be $\delta r= 1.22\lambda /(2 NA) = 370nm$ in the visible range using the Rayleigh criterion.
Because the equipment (and sample) expand when heated, the sample
moves out of the focal plane, and the sample drifts sideways within the field of view upon heating or cooling through the metal-insulator transition. 
To adress this, at each temperature the objective height is swept through the focal length to create a stack of images. The sharpest (i.e. in focus) image in each stack was selected live by finding the highest image gradient using the Tenengrad function. \cite{Liu2016c} The image recording rate during autofocussing was $\sim$12 images/$^o$C. The sideways ($xy$) correction was performed post experiment
using custom image stabilization software, enabling us to 
\textcolor{black}{register each pixel within 50nm.
Although the pixel size is 50$nm$, 
we can only resolve structures within the resolution of 370nm.}
Camera pixel sensitivity was normalized relative to the sapphire substrate in post processing. \cite{instrument_optics}

\textcolor{black}{When VO$_2$ is in a uniform insulating or metallic state, the reflectivity is nearly independent of temperature, even though the resistance changes exponentially with temperature in the insulating state.
The relative difference between the optical reflictivity in the uniform metal (above T=80$^o$C) and uniform insulator (below T=50$^o$C) phases is 20\% in the visible range.   
The raw intensity is recorded in a nearly continuous manner (8-bit grayscale),
with fine temperature steps. These fine steps allowed us to use a Gaussian convolution ($\sigma$=2.5) on the raw optical images over 3 frames to reduce high frequency noise. 
Each pixel's time trace changes in a sufficiently abrupt manner that it is possible to 
identify the moment when each spot on the VO$_2$ film locally undergoes the phase transition.
}
(See  Ref.~\onlinecite{instrument_optics} for full details.) 
This information 
is then used to produce the black and white (metal and insulator) domain maps shown in Fig.~\ref{fig:film_Melissa}.

\subsubsection{Second sample measured by Optics}
\label{sec:SecondSample}
\begin{figure*}
    \begin{center}
    \includegraphics[width=0.9\textwidth,trim={0.4cm 0.0cm 0.8cm 0.0cm},clip]{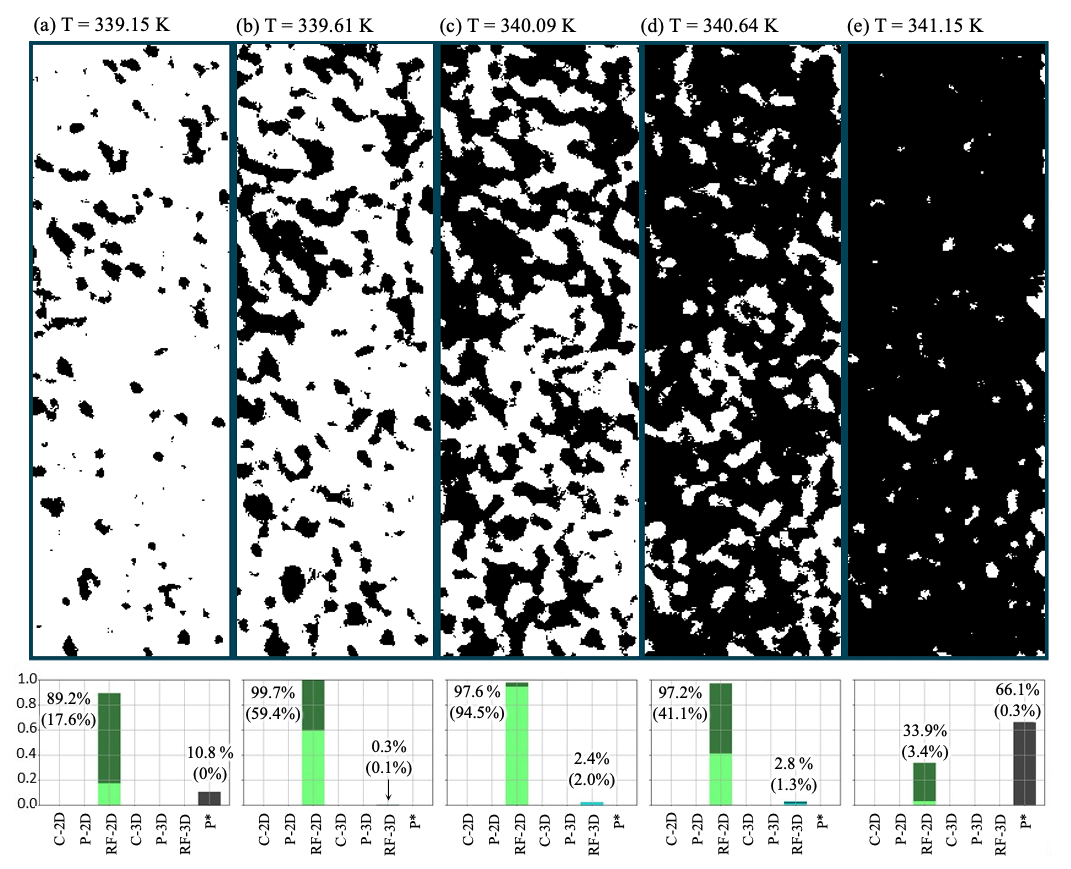}
    \caption{%
    Classification results of our deep learning model applied to optical microscopy images on a second VO$_2$ film, 7.3$\mu m$ wide by 22.7$\mu m$ long.
     White patches are insulating; black patches are metallic.  
        The total percentage of classifications for a particular
        model are reported in the bar charts of panels (a)-(e).
        Classification percentages that fall within $1 \sigma$ of a cluster in the training set are indicated in parentheses.  
        Classifications that fall more than $1 \sigma$
        away from the edge of the corresponding cluster
        in the training set are colored darker in the bar chart. All CNN predictions from optical data  during a temperature ramp up are presented in SI Fig.~\ref{fig:ESPCI-VO2_92_full}.
    }    
    \label{fig:ESPCI-VO2_92}
    \end{center}
\end{figure*}

For consistency we have measured the metal insulator transition in a different VO$_2$ sample using the optical microscope. The main differences with the data presented in the main text can be listed as such:
\begin{itemize}
    \item This dataset was taken during a ramp reversal study.\cite{ramp-reversal-zimmers} The image recording rate during autofocussing was $\sim$2 images/$^o$C producing fewer images taken during the insulator to metal transition than the data shown in the main text.
    \item The gold leads on this sample are separated by only 10 $\mu m$. We apply the same sliding window technique as with the SNIM data to analyze pieces of each image, $100$px $\times$ $100$px at a time,  in each instance returning a classification indicating which Hamiltonian likely produced the pattern formation. Because the optical images in Fig.~\ref{fig:ESPCI-VO2_92} are $ 500 \times 160 $ pixels, this results in $ (500-100+1) \times (160-100+1) = 401 \times 61 $  classifications for each image.
    \item The film is 130nm thick.
    \item The sharpest focus in each stack corresponds to the highest information content, which we identified with a lossless Tiff format using a standard Lempel-Ziv-Welch compression algorithm.\cite{LZW,LZW2}. The gradient method using the Tenengrad function was found to give identical results but is faster computationally.
\end{itemize}

\begin{figure*}
    \begin{center}
    \includegraphics[width=.49\textwidth]{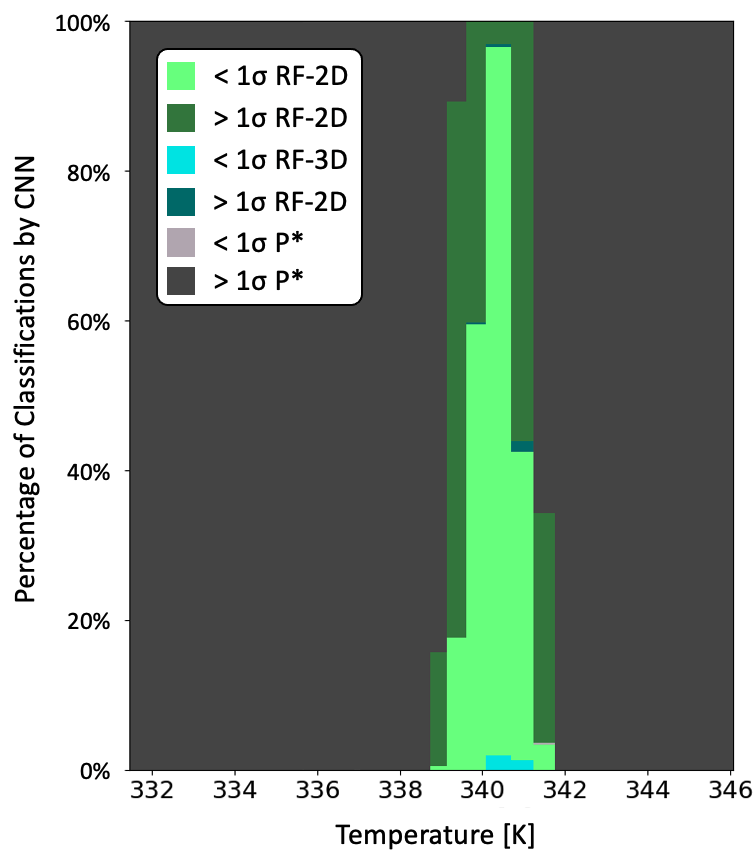}%
    \caption{%
        All CNN predictions from optical data during a temperature ramp up of data presented in Fig.~\ref{fig:ESPCI-VO2_92}.
        Darker colors denote classifications that are more than 1 standard deviation from the identified training set.
    \label{fig:ESPCI-VO2_92_full}}
    \end{center}
\end{figure*}

The ML results presented in Figures~\ref{fig:ESPCI-VO2_92} and ~\ref{fig:ESPCI-VO2_92_full}  agree with the results presented in the main part of the paper, Figures~\ref{fig:film_Melissa} and ~\ref{fig:film_full_Melissa}.

\section{Clean Ising Models}
We simulated the clean Ising model using a mixture of Monte Carlo updates including a parallel checkerboard Metropolis update,\cite{metropolis}
as well as the Wolff algorithm.\cite{wolff} 
The system was thermalized for $100,000$ steps of parallel checkerboard Metropolis interleaved with $10,000$ Wolff updates.
After thermalization, spin configurations were saved after $10$ parallel checkerboard Metropolis and $10$ Wolff updates.
We simulate systems of size $100 \times 100 \times 100$, with open boundary conditions (OBC) in the z direction, and periodic boundary conditions (PBC) in the x and y directions.   After thermalization, we save spin configurations from the two open surfaces, as well as from 3 paralell 2D slices taken from the interior.  All 5 of the 2D spin configurations thus generated are equally spaced.  We use a sparse covering in order to minimize correlations between the 2D spin configurations thus obtained.  We also save 2D spin configurations from the two orthogonal directions. Since there are periodic boundary conditions in these directions, we can only generate 4 equally spaced configurations perpendicular to the x direction, and 4 perpendicular to the y direction.  Thus, a total of $5+4+4 = 13$ 2D spin configurations are saved from each measurement step of the simulations.  This allows us to cover a range of surface effects that can happen in experimental data taken at free surfaces of large bulk samples ({\em i.e.} small samples well as large samples viewed on a small window), and save much computational time.  If we only use open surfaces of, say, large systems that we window down to a $100 \times 100$  FOV, we would need to simulate larger system sizes, typically $400 \times 400 \times 400$ with OBC, but only save the 6 free surfaces.  This method would take $4\times 4 \times 4 = 64$ times longer to simulate, plus we would need to run $13/6 \approx 2$ times longer to generate as many configurations.  Our method with mixed boundary conditions, and saving some interior and exterior slices, allows us to cover a range of surface effects that can happen at a real surface, where the effective  fluctuations can be affected not just by the free surface but also by surface reconstruction effects, but takes $4 \times 4 \times 4 \times 13 /6 \approx 138$ times less computational time to generate the same number of spin configurations for training.  One can see from Fig.~\ref{fig:dist_out} that the DL classifier places configurations from both free surfaces and interior slices into the same cluster, rather than producing a distinct cluster for each case.

\section{Simulating multiscale structures}


In order to train the DL classifier to recognize multiscale pattern formation in the above models, we simulate them near their transitions and feed these sets of spin configurations into a deep neural network framework that is designed for pattern recognition in images.
We generate simulation results from $100\times100$ (2D) and $100\times100\times100$ (3D) lattice size systems.

The parameters from Table~\ref{tab:sim_params} are used to generate 8000 images for each model near it transition.  
The exception is P$^*$, which is a set of images not
near a transition, 
for which we generate 16000 images
because it covers a larger region of parameter space.  
Other than P$^*$, the models each have a second order phase transition,
at which the spin configurations display power law behavior,
with structure on all length scales up to the correlation length, which
diverges at the critical point.  We focus on models near a second
order phase transition, because the multiscale behavior inherent to
criticality is capable of driving the multiscale pattern formation
often seen in spatially resolved experiments on quantum materials.\cite{dagotto,phillabaum-2012,comin-ndnio,shuo-prl,kohsaka-science,qazilbash-science}

\begin{figure*}
    \begin{center}
        \includegraphics[width=0.7\textwidth]{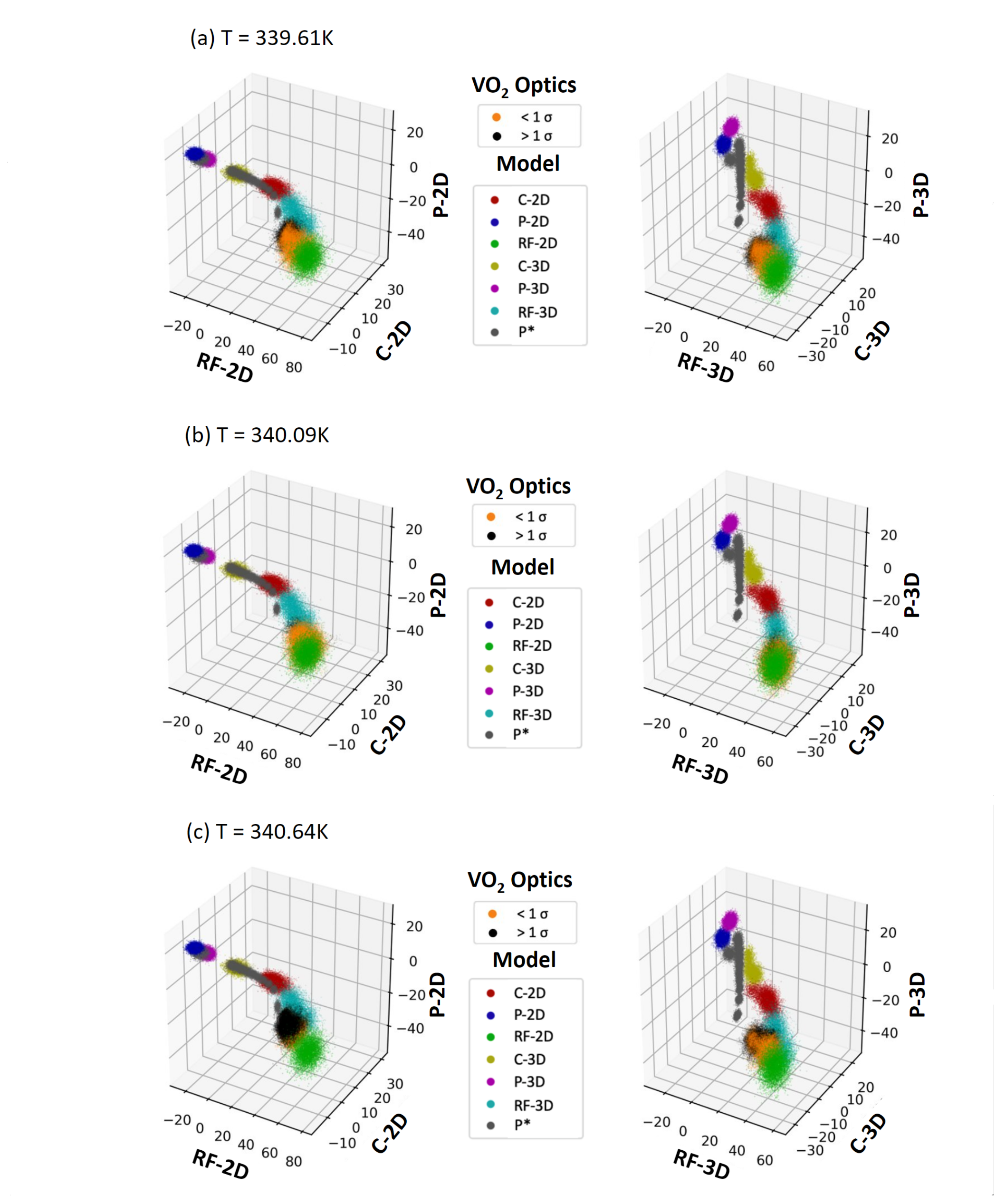}
    \caption{%
    \label{fig:MQ_soft_d_optics_rect}
            Distribution of relative weights of each class in the last fully connected layer, for the VO$_2$ optical data (rectangular second sample presented in the supplementary text Sec. \ref{sec:SecondSample} and Fig. \ref{fig:ESPCI-VO2_92}), superimposed on the distribution for the training sets shown in Fig.~\ref{fig:dist_out}.
        Results for the VO$_2$ data that are within one $7$-dimensional standard deviation of a training set are indicated by orange  dots.
       Results for the VO$_2$ data that are farther away are indicated by black dots.   
    }
    \end{center}
\end{figure*}

\end{document}